\documentclass[pre, reprint, superscriptaddress, amsmath, amssymb, aps]{revtex4-2}

\usepackage{graphicx}
\usepackage{dcolumn}
\usepackage{bm}
\usepackage[protrusion=true,expansion=true]{microtype}
\usepackage[normalem]{ulem}
\usepackage{soul}
\usepackage{xcolor}
\usepackage{kotex}
\usepackage{hyperref}
\usepackage{cleveref}
\usepackage{bibunits}

\defaultbibliographystyle{apsrev4-2}
\defaultbibliography{ref}

\usepackage{comment}

\def\min{\textrm{min}}
\def\max{\textrm{max}}

\DeclareMathOperator{\sech}{sech}

\begin{document}

\begin{bibunit}

\preprint{APS/123-QED}

\title{Paths to synchronization in the Kuramoto model with inertia}

\author{Cook Hyun Kim}
\affiliation{CCSS, KI for Grid Modernization,
Korea Institute of Energy Technology, Naju, Jeonnam 58330, Korea}

\author{Yeomoon Kim}
\affiliation{CCSS, KI for Grid Modernization,
Korea Institute of Energy Technology, Naju, Jeonnam 58330, Korea}

\author{Stefano Boccaletti}
\affiliation{Sine-Europe Complexity Science Center, North University of China, 3 Xueyuan, Taiyuan, Shanxi, 030051, China}
\affiliation{Research Institute of Interdisciplinary Intelligent Science, Ningbo University of Technology, 201 Fenghua, Ningbo, Zhejiang, 315211, China}
\affiliation{CNR, Institute of Complex Systems, Madonna del Piano 10, Sesto Fiorentino, Firenze, 50019, Italy}

\author{B. Kahng}
\email{bkahng@kentech.ac.kr}
\affiliation{CCSS, KI for Grid Modernization,
Korea Institute of Energy Technology, Naju, Jeonnam 58330, Korea}

\date{\today}

\begin{abstract}
Synchronization is ubiquitous across natural and synthetic systems, yet most prior studies focus on the inertia-free Kuramoto model and do so at the macroscopic level. 
In this study, we instead investigate the inertial Kuramoto model and analyze the kinetics of individual synchronized clusters that emerge in the underdamped dynamics, driven by the interactions among multiple synchronized clusters with different frequencies. Specifically, we explore two forms of intrinsic frequency distribution---unimodal Gaussian and multimodal uniform---and show that they give rise to qualitatively different synchronized clusters: a hierarchical organization for the Gaussian distribution and a homogeneous organization for the uniform distribution. This contrast leads to qualitatively different behaviors of the order parameter: for the Gaussian distribution, it increases smoothly with increasing coupling strength, while for the uniform distribution, it grows through a series of discrete jumps that trace out the size of the Devil's staircase (DS). 
By resolving the kinetics at the cluster level, we further find that the route to synchronization also depends on the distribution type: with a Gaussian distribution, a single dominant cluster forms and gradually entrains the remaining oscillators, whereas with a uniform distribution, synchronization proceeds via successive cluster mergers initiated from peripheral seeds associated with the high-frequency periphery.
Taken together, these findings provide a new perspective on collective synchronization dynamics in inertial complex systems.
\end{abstract}

\maketitle


\section*{Introduction}

Synchronization phenomena are ubiquitous in biological~\cite{buck1938synchronous, buck1966biology, buck1976synchronous, buck1988synchronous}, neuronal~\cite{cumin2007generalising, breakspear2010generative, bick2020understanding}, and engineered systems~\cite{grainger1999power, anvari2020introduction, guo2021overviews, witthaut2022collective, forrester2015arrays}. The collective dynamics of coupled oscillators have been studied extensively across complex network topologies, where the interplay between structure and dynamics produces a wide range of synchronization phenomena~\cite{boccaletti2018synchronization}. While many theoretical frameworks assume that the instantaneous adaptation of oscillator frequencies to coupling forces~\cite{kuramoto1975international, kuramoto1984chemical, strogatz2000kuramoto, acebron2005kuramoto, strogatz2018nonlinear} is the main factor, real physical systems are also subject to inertial effects~\cite{tanaka1997first, tanaka1997self, gao2018self}, which impose finite response times and thereby fundamentally modify the synchronization dynamics~\cite{olmi2014hysteretic, gao2021synchronized, kim2025cluster, kim2026predicting, yi2026phase}. These effects are captured by the second-order Kuramoto model (2nd KM), which is written as
\begin{align}
m\ddot{\theta}_i + \gamma \dot{\theta}_i = \omega_i + \frac{K}{N}\sum_{j=1}^{N} \sin(\theta_j - \theta_i),
\label{eq:2nd_kuramoto}
\end{align}
where $\theta_i$ is the phase of the oscillator $i$, and $\dot{\theta}_i$ and $\ddot{\theta}_i$ are its angular velocity and acceleration, respectively. 
Here $m$ and $\gamma$ are the inertia and the damping coefficient, respectively. The intrinsic frequency $\omega_i$ is drawn from a distribution $g(\omega)$, $K$ is the coupling strength, and $N$ is the number of oscillators in an all to-all topology. The model describes systems ranging from Josephson junction arrays~\cite{josephson1962possible, anderson1963probable, josephson1974discovery} to, in modified form, electric power grids~\cite{park2025optimal, lee2024reinforcement, park2026ai}. Synchronization is measured by the order parameter $R$, defined through $R e^{\mathrm{i}\Theta} = \frac{1}{N}\sum_{i=1}^{N} e^{\mathrm{i}\theta_i}$: for $K < K_c$ the system is incoherent ($R=0$), whereas for $K > K_c$ it becomes coherent ($R>0$).

For a unimodal $g(\omega)$ such as a Gaussian, the transition in the 2nd KM is discontinuous, unlike the continuous transition of the inertia-free first-order Kuramoto model (1st KM). The 2nd KM exhibits hierarchical clustering: whereas the 1st KM forms a single synchronized cluster, the 2nd KM supports multiple coexisting clusters with distinct angular velocities. A primary cluster forms first and acts as an attractor~\cite{tanaka1997first, tanaka1997self, gao2018self}; secondary and higher-order clusters then trace elliptical orbits around it, making the order parameter oscillate~\cite{gao2018self}. Frequency resonances lock the clusters into a Devil's staircase (DS) of rational rotational-frequency ratios~\cite{kim2025cluster}. All of this stems from inertia, which enables clusters to preserve distinct rotational frequencies instead of collapsing onto a single frequency-locked state.

While the Gaussian case has been extensively investigated, other forms of $g(\omega)$ remain largely unexplored. Even for the 1st KM, Gaussian and uniform distributions lead to qualitatively different transition behaviors: the Gaussian distribution results in a continuous transition, whereas the uniform distribution gives rise to a mixed-order transition, with $R - R_c \sim (K-K_c)^{2/3}$~\cite{pazo2005thermodynamic}. After the transition point, however, both distributions display a smooth increase of $R$ toward complete synchronization, forming a single macroscopic cluster. This similar behavior suggests that the primary role of the different distribution types is to modify the driving mechanism of the transition rather than the subsequent route to synchronization. Note that in the 1st KM with a uniform distribution, the width of $g(\omega)$ merely rescales $K_c$, leaving the transition order and critical exponents unchanged~\cite{song2020effective}.

The uniform distribution in the 2nd KM is of more than theoretical interest. In alternating current power grids, the swing equation, which describes the stabilization of frequency deviations, shares the dynamical structure of the 2nd KM. The power injected at each bus plays the role of the intrinsic frequency. As the fraction of renewable-energy generators increases, the injected-power distribution shifts from a Gaussian-like shape toward more uniform shapes~\cite{anvari2020introduction, guo2021overviews, witthaut2022collective}. Thus, synchronization under uniform distributions bears directly on the stability of modern power grids.

In this situation, it is interesting to consider how the uniformity of the $g(\omega)$ distribution in the 2nd KM affects the synchronization transition; whether it also controls only the character of the transition or reshapes the entire collective dynamics. An earlier work recognized that inertia modifies the transition~\cite{tanaka1997self}, but how the distribution shape controls cluster's formation and organization has not been addressed. We find that when inertia is present, the shape of the intrinsic frequency distribution influences the entire collective dynamics; not only the nature of the transition but also additional phenomena such as pronounced multistability, characterized by the coexistence of numerous stable cluster states~\cite{pisarchik2022multistability}.

\begin{figure*}[!t]
\centering
\includegraphics[width=1.0\linewidth]{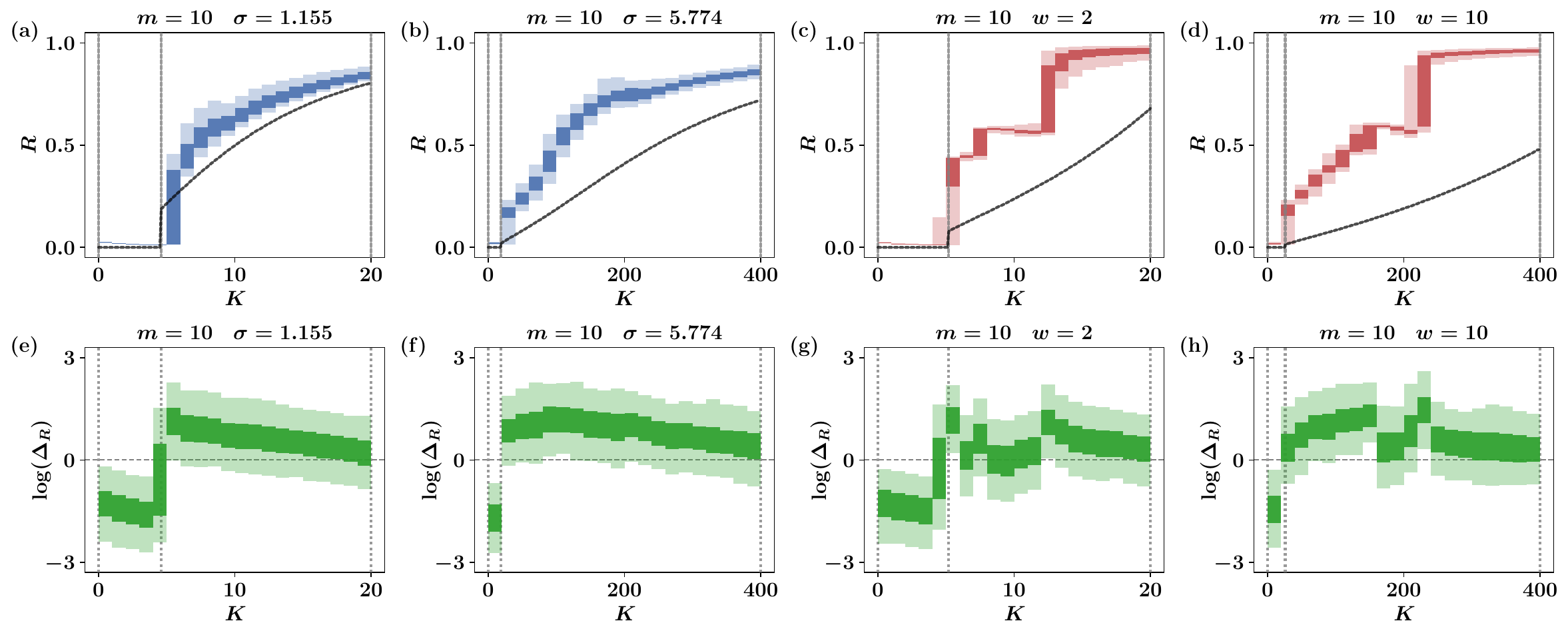}
\caption{
{\bf Shape-dependent growth of the order parameter and its susceptibility.}
(a--d) Order parameter $R$ versus coupling strength $K$ for $m=10$ and Gaussian $\sigma=1.155$ (a), Gaussian $\sigma=5.774$ (b), uniform $w=2$ (c), and uniform $w=10$ (d); the black curve and the gray vertical line represent the solution of the SCE~\eqref{eq:self_consistency} and a transition point, respectively; and the shaded area represents the variability over $256$ realizations: the darker band corresponds to the interquartile range, while the lighter band shows the $5$–$95\%$ interval.
(e--h) The corresponding susceptibility $\log\Delta R$ [Eq.~\eqref{eq:deltaR}], where the dark and light bands again indicate the interquartile and $5$--$95\%$ ranges over realizations; the locations of its peaks identify the coupling strengths at which the system is maximally sensitive.
The specific form of $g(\omega)$ dictates how $R$ grows: for a Gaussian distribution, $R$ increases smoothly and exhibits a single susceptibility peak at the transition point; in contrast, for a uniform distribution, $R$ rises through a sequence of discrete jumps, each associated with its own susceptibility peak, as though every jump represented a separate transition. The global profile of the distribution thus determines whether the behavior is continuous or stepwise, while the width $w$ of the uniform distribution simply stretches the range of $K$ over which this pattern is observed [SM].
}
\label{fig:rev1}
\end{figure*}

\begin{figure*}[!t]
\centering
\includegraphics[width=1.0\linewidth]{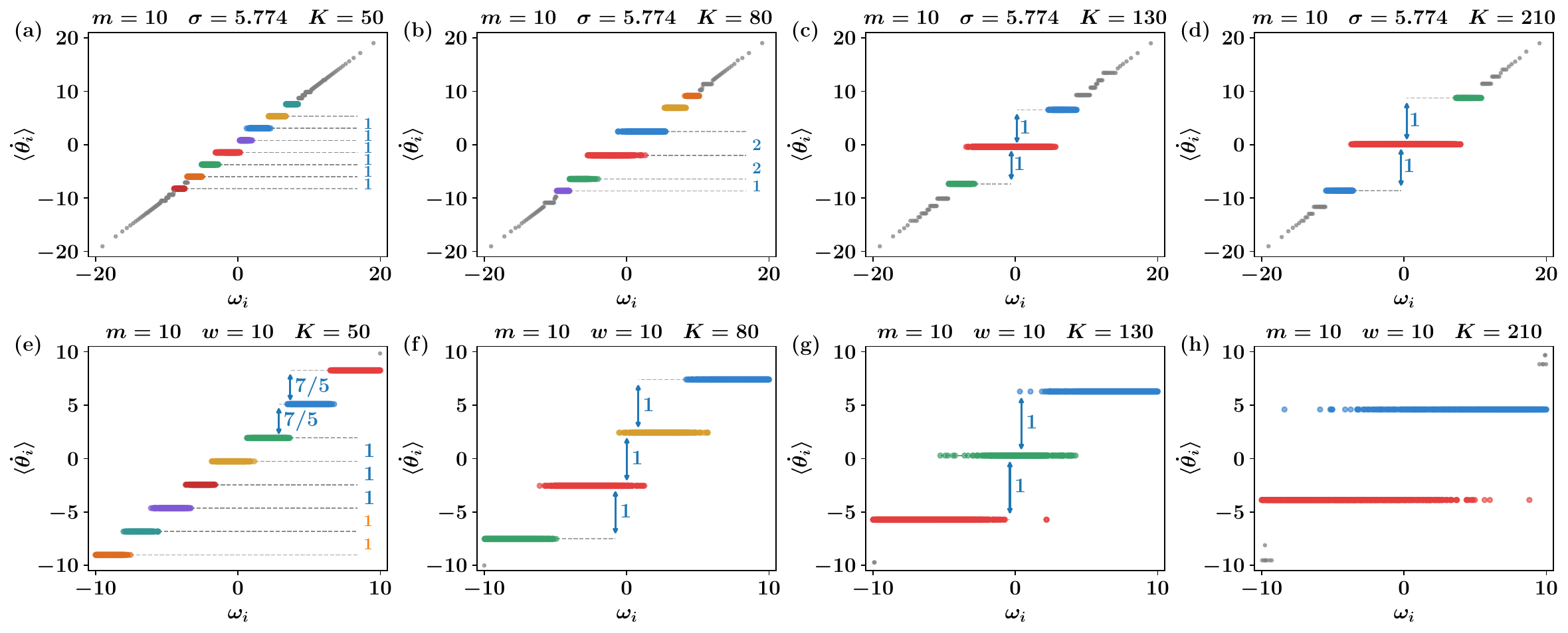}
\caption{
{\bf Hierarchical versus homogeneous cluster organization.}
Time-averaged angular velocity $\langle\dot\theta_i\rangle$ versus intrinsic frequency $\omega_i$ for a single realization ($m=10$). (a--d) Gaussian with $\sigma=5.774$ and (e--h) uniform with $w=10$ at $K=50$, $80$, $130$, and $210$. Colored plateaus indicate groups that are synchronized, and arrows denote the rational ratios associated with the angular-velocity gaps between neighboring groups. The Gaussian distribution produces a hierarchical Devil's staircase centered around one dominant cluster, separated by uneven gaps with ratios like $2\!:\!1$. In contrast, the uniform distribution yields clusters of comparable size, spaced by uniform gaps ($1\!:\!1\!:\!1$), reflecting a homogeneous form of the Devil's staircase.
}
\label{fig:rev2}
\end{figure*}

\begin{figure*}[!t]
\centering
\includegraphics[width=1.0\linewidth]{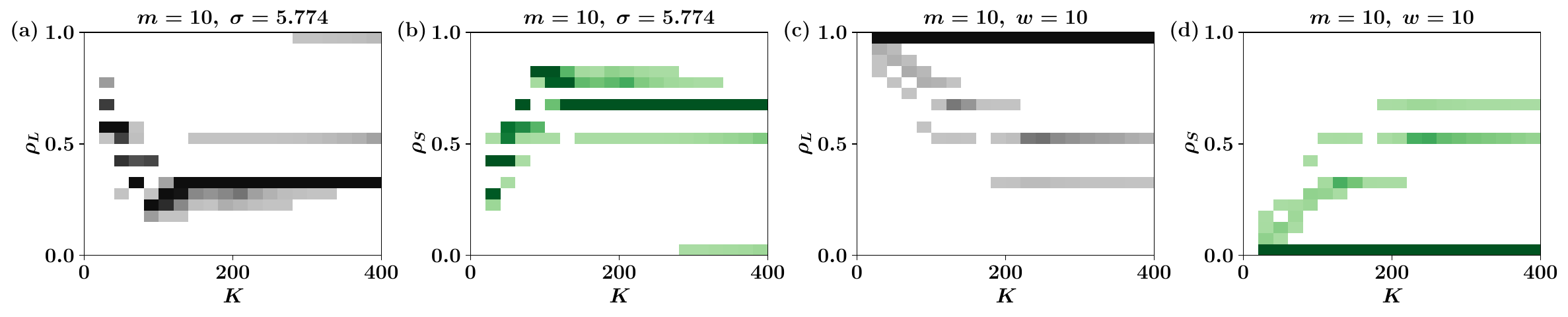}
\caption{
{\bf Cluster number fractions.}
Number fractions $\rho_L$ and $\rho_S$ of large and small clusters ($\rho_L+\rho_S=1$) versus coupling strength $K$ for $m=10$.
(a,b)~Gaussian with $\sigma=5.774$ and (c,d)~uniform with $w=10$, showing $\rho_L$ and $\rho_S$ in turn.
Large clusters form a minority in the Gaussian case ($\rho_L$ low, $\rho_S$ high), with the single major cluster being outnumbered by numerous smaller satellite clusters, whereas in the uniform case the system is composed almost exclusively of large clusters ($\rho_L\to1$, $\rho_S\to0$).
}
\label{fig:rev5}
\end{figure*}

\begin{figure*}[!t]
\centering
\includegraphics[width=1.0\linewidth]{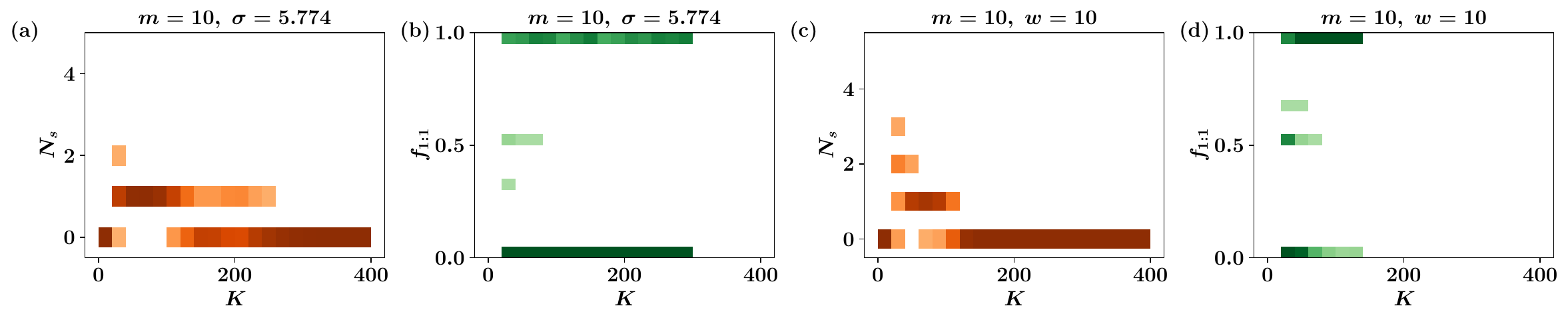}
\caption{
{\bf Shape-dependent Devil's staircase ratios.}
Modal number of staircases $N_s$ (a,c) and equal-gap fraction $f_{1:1}$ (b,d) versus coupling strength $K$ for $m=10$: (a,b) Gaussian $\sigma=5.774$ and (c,d) uniform $w=10$.
Each quantity is shown as a shaded histogram over the ensemble of realizations at each $K$ (darker represents more frequent).
As $K$ increases, $N_s$ decreases because clusters coalesce, approaching complete locking (a,c).
The distribution shape determines which ratios emerge (b,d): $f_{1:1}$ is large for the uniform case, where clusters of comparable size generate equal ($1\!:\!1\!:\!\cdots$) spacings, but it is small for the Gaussian, where a dominant central cluster imposes hierarchical ratios such as $2\!:\!1$.
The accuracy of the locking is characterized in the SM [Fig.~\ref{fig:figS_RevS9}].
}
\label{fig:rev12}
\end{figure*}

We find that the shape of the $g(\omega)$ distribution controls synchronization dynamics in the following ways:
\begin{itemize}
\item[(i)] The shape controls how the order parameter increases: it varies smoothly for the Gaussian distribution, showing a single susceptibility peak at the transition, whereas for the uniform distribution it grows via discrete jumps. Each jump corresponds to a reorganization of clusters, each associated with its own peak. The single mean-field self-consistency equation correctly identifies the critical coupling \(K_c\), but it breaks down beyond the transition, where inertia-driven multi-cluster dynamics dominate [Fig.~\ref{fig:rev1}].
\item[(ii)] The shape determines how the clusters arrange themselves. Because the uniform shape has no central peak, it generates clusters of similar size that have evenly spaced $1\!:\!1$ differences in their mean angular velocities, leading to a homogeneous DS. This stands in contrast to the irregular, hierarchical ratios observed in the Gaussian case [Figs.~\ref{fig:rev2}, \ref{fig:rev5}, and \ref{fig:rev12}].
\item[(iii)] The uniform shape determines how the giant cluster forms: it is no longer assembled through central entrainment, but rather through the merger of peripheral clusters, so its seeds come to lie at the high-frequency periphery rather than the center [Figs.~\ref{fig:rev4}, \ref{fig:rev6}, \ref{fig:rev13}, \ref{fig:rev9}, and \ref{fig:rev10}].
\item[(iv)] The shape thus determines stability: by constraining the assembly pathway, the entrainment-driven and merger-driven mechanisms remain robust, but in complementary coupling regimes [Fig.~\ref{fig:rev11}]. A detailed account of the distinct contributions of the variance and the energy-based capture mechanism is provided in the SM [Secs.~\ref{sec:melnikov_failure}--\ref{sec:initial_energy}].
\end{itemize}

\section*{Main Results}

We set a Gaussian and a uniform distribution that share the same variance ($\sigma = w/\sqrt{3}$), so that $\sigma = 1.155$ and $5.774$ correspond to $w=2$ and $10$, respectively. Note that an intrinsic frequency $\omega_i$ is selected in the range $-w < \omega_i < w$. Throughout the main text, we set $\gamma = 1$ and $m = 10$. We take $m = 10$ and a large variance pair ($\sigma = 5.774$, $w = 10$) for our main analysis because the multi-cluster, interaction-driven dynamics studied here emerge when the inertia and the variance are large enough that multiple clusters coexist and interact. When either is small, too few clusters coexist to interact: the system relaxes into a single giant cluster, well described by the macroscopic order parameter and its single-cluster self-consistency equation, and inertia merely changes the order of the transition, as already established~\cite{tanaka1997first, tanaka1997self}. Instead, when the inertia and the variance are large enough, inertial effects become dominant and reorganize the full collective dynamics, not just at the transition point.

\subsection*{Order Parameter Behaviors for the Two $g(\omega)$ Distributions}

We consider the transition across coupling strengths $K$, taking initial phases $\theta_i$ uniform on $[-\pi, \pi]$ and angular velocities $\dot{\theta}_i = \omega_i/\gamma$. The black curve in Fig.~\ref{fig:rev1}(a--d) is the solution of the self-consistency equation (SCE)~\cite{tanaka1997first, tanaka1997self},
\begin{align}
R = \int_{\omega_c - \omega_M}^{\omega_c + \omega_M} d\omega \, g(\omega) \sqrt{1 - \left(\dfrac{\omega - \omega_c}{KR}\right)^{2}},
\label{eq:self_consistency}
\end{align}
where $\omega_c$ is the mean frequency of the cluster. The entrainment boundary $\omega_{M} = (4\gamma/\pi)\sqrt{KR/m}+\alpha\sqrt{1/(KRm^{3})}$, with $\alpha \approx -0.3056$, comprises a leading term derived from the Melnikov method~\cite{melnikov1963stability, guckenheimer2013nonlinear, gao2018self} and an empirical subleading correction taken from Ref.~\cite{gao2018self} [SM, Sec.~\ref{sec:melnikov_derivation}].

The form of $g(\omega)$ determines how $R$ grows, and this growth leaves a matching imprint on the susceptibility. We define the ensemble-averaged ratio of the final to the initial deviations of $R$ from their means,
\begin{align}
\Delta R = \left\langle \frac{\lvert R^{\mathrm{final}} - \langle R^{\mathrm{final}}\rangle\rvert}{\lvert R^{\mathrm{initial}} - \langle R^{\mathrm{initial}}\rangle\rvert} \right\rangle,
\label{eq:deltaR}
\end{align}
where $R^{\mathrm{initial}}$ and $R^{\mathrm{final}}$ are the order parameter of a single realization, time-averaged over the initial interval $t\in[0,4]$ and the final interval $t\in[9600,t_f]$, respectively. The initial interval captures the small finite-size fluctuations of the incoherent state, so that $\Delta R$ measures how strongly these initial differences are amplified into the final state. $\langle R^{\mathrm{initial}}\rangle$ and $\langle R^{\mathrm{final}}\rangle$ are their means over the ensemble of realizations, and the outer $\langle\,\cdot\,\rangle$ likewise averages the ratio over the ensemble. Here $R$ plays the role of magnetization and $\Delta R$ that of susceptibility, so a peak in $\log\Delta R$ [panels (e--h)] marks a critical coupling at which the system is critically sensitive.

For the Gaussian distribution with $\sigma = 1.155$ [Fig.~\ref{fig:rev1}(a)], $R$ increases steeply at the transition point and then continues to grow smoothly until full synchronization is achieved, while $\log\Delta R$ displays a single peak there and then declines [Fig.~\ref{fig:rev1}(e)]—a signature of a standard transition governed by a single critical point. In contrast, for the uniform distribution, the order parameter increases via a succession of discrete jumps, with plateaus interrupted by abrupt rises in $R$, already visible at small width [$w = 2$, Fig.~\ref{fig:rev1}(c)]; in this case, $\log\Delta R$ does not simply decay but instead shows a new peak at each jump [Fig.~\ref{fig:rev1}(g)], as if every jump marked an independent transition. This difference in behavior arises from the distinct shapes of the $g(\omega)$ distributions, specifically whether $g(\omega)$ possesses a central peak or lacks one.

\subsection*{Breakdown of the Self-Consistency Equation}

The numerical behavior of $R$ and $\Delta R$, compared with the corresponding SCE solutions in Fig.~\ref{fig:rev1}, indicates that the SCE solution ceases to be valid beyond the transition point. For the Gaussian distribution with a small standard deviation $\sigma=1.155$ in Fig.~\ref{fig:rev1}(a), the order parameter is expected to approximately follow the mean-field prediction; however, the numerical data already diverge from the SCE solution (solid curve) in the vicinity of $K_c$. In contrast, when $\sigma=5.774$ is relatively large, as in Fig.~\ref{fig:rev1}(b), and likewise for the uniform distribution, the numerical results exhibit a clear and substantial deviation from the SCE prediction. These discrepancies arise from inertia-driven multi-cluster dynamics and the interactions among clusters, which reshape the overall collective behavior. Each jump in $R$ and each peak in $\log\Delta R$ corresponds to a discrete reorganization of the cluster structure. As a consequence, the macroscopic order parameter alone is insufficient to describe the dynamics. The transition must instead be analyzed at the mesoscopic cluster level, accounting for the number of clusters, their sizes, and the frequency separations between them. We stress that the specific sequence of cluster rearrangements is sensitive to the particular form of $g(\omega)$. In what follows, we focus on this cluster-level analysis.

\subsection*{Hierarchical versus Homogeneous Cluster Organization}

Fig.~\ref{fig:rev2} shows the time-averaged angular velocity $\langle\dot\theta_i\rangle$ versus the intrinsic frequency $\omega_i$ for a single realization. Colored plateaus mark synchronized clusters. Arrows between the two clusters indicate that the gaps between neighboring angular velocities follow the rational ratios.

The two $g(\omega)$ distributions organize their clusters in qualitatively different ways. For the Gaussian case [Fig.~\ref{fig:rev2}(a--d)], synchronized clusters are built in the following way: one dominant cluster is set up at the center $\langle\dot\theta_i\rangle = 0$, and the other subsidiary clusters are arranged hierarchically around it. Thus, neighboring gaps can have unequal ratios such as $2\!:\!1$ [Fig.~\ref{fig:rev2}(b)]. For the uniform case [Fig.~\ref{fig:rev2}(e--h)], the clusters are of comparable size across the frequency range, with nearly equal gaps in a $1\!:\!1\!:\!1$ pattern [Fig.~\ref{fig:rev2}(e)]. We refer to this formation as a homogeneous cluster organization. As $K$ increases, clusters merge, and the staircase simplifies in both cases until only two or three large clusters remain in a steady state. In short, for the Gaussian distribution, the synchronized clusters emerge hierarchically, whereas for the uniform distribution, they organize in a homogeneous manner.

At the final time $t_f$, the cluster profiles have become stationary. We then classify the clusters into large and small ones: after determining the size of the largest cluster, $s_{\rm max}$, we define the characteristic size as $s_c = s_{\rm max}/2$. Clusters with sizes exceeding $s_c$ are labeled as large, while the rest are categorized as small. The populations of these clusters, aggregated over different ensembles, are displayed in Fig.~\ref{fig:rev5}. We interpret the collection of large clusters as macroscopic-scale clusters. The number fractions of large and small clusters are denoted by $\rho_L$ and $\rho_S$, respectively, with $\rho_L + \rho_S = 1$.

Fig.~\ref{fig:rev2} shows that for the Gaussian case, the large clusters are minor while the small clusters are abundant. [Fig.~\ref{fig:rev5}(a,b)] demonstrates that, in the Gaussian case, one dominant cluster is accompanied by numerous much smaller satellite clusters, whereas for the uniform case, the clusters are almost exclusively large ($\rho_L\to1$, $\rho_S\to0$) [Fig.~\ref{fig:rev5}(c,d)]. This contrast in size holds over a wide range of inertia and variance values, serving as a robust indicator of the distribution's shape [Sec.~\ref{sec:RS5rho}].

The form of $g(\omega)$ also shapes the DS they lock into. For both distributions, the angular-velocity gaps settle onto precise rational ratios---the defining feature of the staircase. We measure this precision by the normalized gap variance $\mathcal{V}_{\Delta\dot\theta}$. Together with the staircase participation $f_{\mathrm{stair}}$, we confirm that the locking is precise and that almost every cluster takes part, regardless of the distribution. Because these two properties establish the staircase itself rather than distinguish the two shapes, their definitions and data are deferred to the SM [Sec.~\ref{sec:RS6}, Fig.~\ref{fig:figS_RevS9}].

The form of $g(\omega)$ also determines the ratios that arise, quantified by the equal-gap fraction $f_{1:1}$---the proportion of staircases whose integer ratios $p_k$ are all equal (equivalently, whose gaps are all identical), forming a $1\!:\!1\!:\!\cdots\!:\!1$ pattern [Fig.~\ref{fig:rev12}(b,d)]. It is large for the uniform case, where clusters of comparable size create evenly spaced gaps, but small for the Gaussian case, where a dominant central cluster imposes hierarchical ratios like $2\!:\!1$. As $K$ grows, the most common number of staircases $N_s$ declines as clusters coalesce [Fig.~\ref{fig:rev12}(a,c)], eventually disappearing as the system approaches full locking. This difference between the two $g(\omega)$ reflects a robust signature of the distribution's shape, rather than any specific choice of inertia or variance [Sec.~\ref{sec:RS6}].

\begin{figure*}[!t]
\centering
\includegraphics[width=1.0\linewidth]{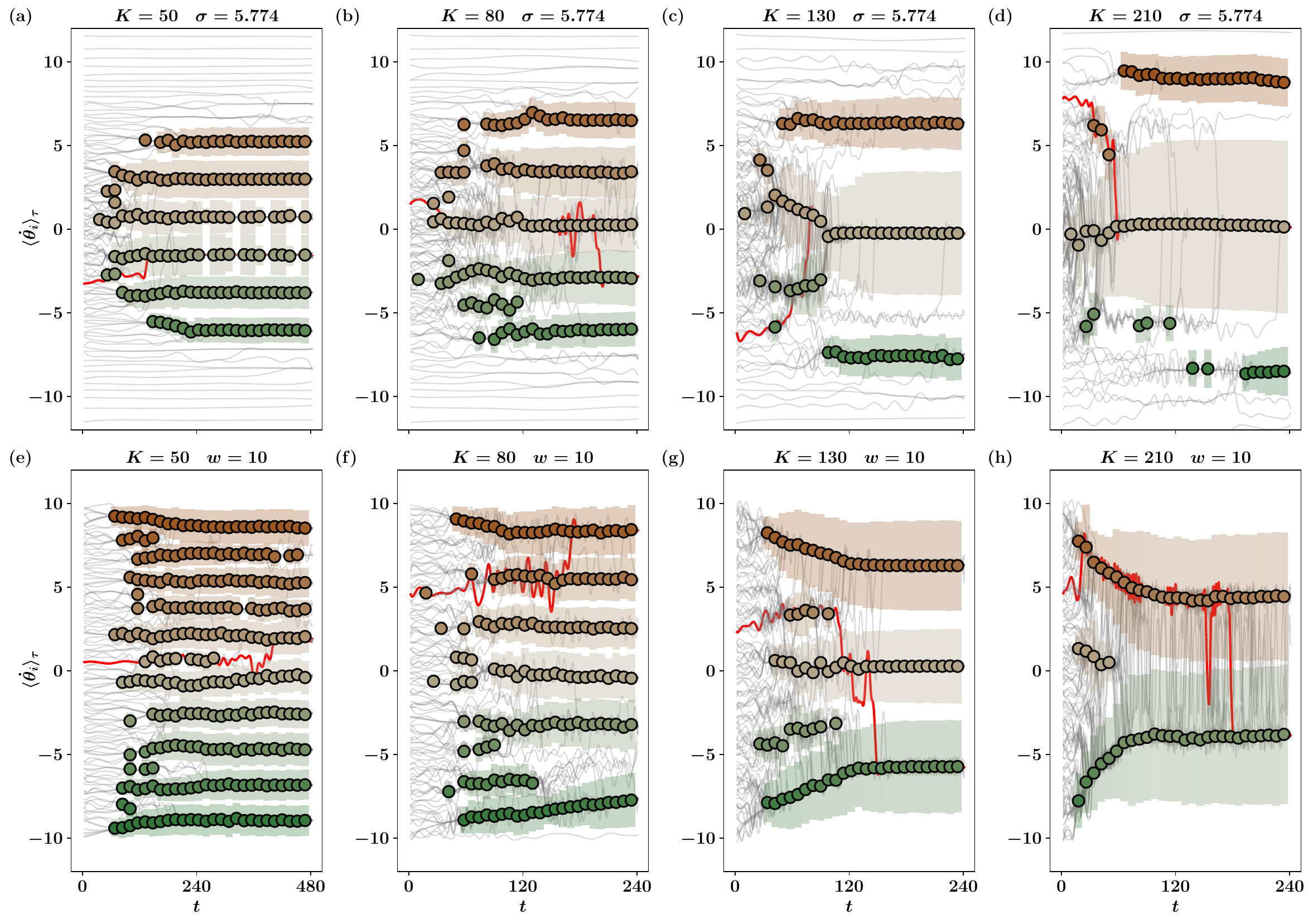}
\caption{
{\bf Cluster formation dynamics.}
Plot of the average angular velocity $\langle\dot\theta_i(t)\rangle_\tau$ as a function of time. Here, $\langle \cdot \rangle_\tau$ means the average over time interval $[t,t+\tau]$ for visibility, where $\tau$ is a short averaging window, taken as $\tau=8$ at $K=50$ and $\tau=4$ at all larger $K$. (a--d) For the Gaussian cases with $\sigma=5.774$ and (e--h) for the uniform case with $w=10$, at $K=50$, $80$, $130$, and $210$.
The two shapes generate their clusters through different mechanisms. In the Gaussian case, a single dominant cluster first appears at the center and then extends outward by entraining nearby oscillators, with additional smaller clusters forming later. In contrast, for the uniform case, multiple clusters of comparable size emerge almost simultaneously across the frequency range and subsequently coalesce through merging—a homogeneous mode of formation. Gray thin curves show the trajectories of $128$ oscillators regularly selected from the full ensemble of $N=1024$. The red curves indicate those oscillators that switch to different clusters; their intermittent paths demonstrate that oscillators are continuously exchanged between clusters and that inter-cluster coupling disturbs their motion. The shaded background regions around each cluster depict the Melnikov region, bounded by $\pm\omega_M$.
}
\label{fig:rev4}
\end{figure*}

\begin{figure}[!t]
\centering
\includegraphics[width=1.0\linewidth]{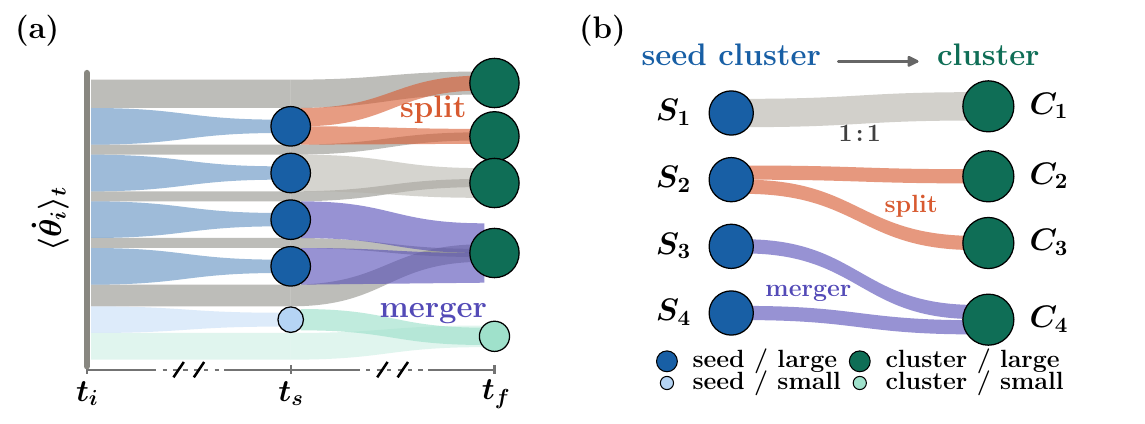}
\caption{
{\bf Birth and development of clusters.}
(a)~Schematic of cluster evolution over time; large circles denote large clusters, and $\langle\dot\theta_i\rangle_\tau$ is the mean angular velocity defined in Fig.~\ref{fig:rev4}. The clusters identified at $t_s$ and $t_f$ are termed seed clusters and clusters, respectively, and $t_s$ is the earliest time at which the number of large seed clusters exceeds that of large clusters. Large and small clusters, and large and small seed clusters, are distinguished by the size thresholds defined in Sec.~\ref{sec:cluster_id}.
(b)~A large seed cluster maps onto a cluster at $t_f$ when the two share at least $\sqrt{N}$ oscillators, giving a many-to-many correspondence: a seed cluster typically maps onto a single large cluster, but it may map onto two, or two seed clusters may map onto one. This mapping underlies the transient capture probability $P_{\mathrm{ts}}$.
}
\label{fig:rev6}
\end{figure}

\begin{figure*}[!t]
\centering
\includegraphics[width=1.0\linewidth]{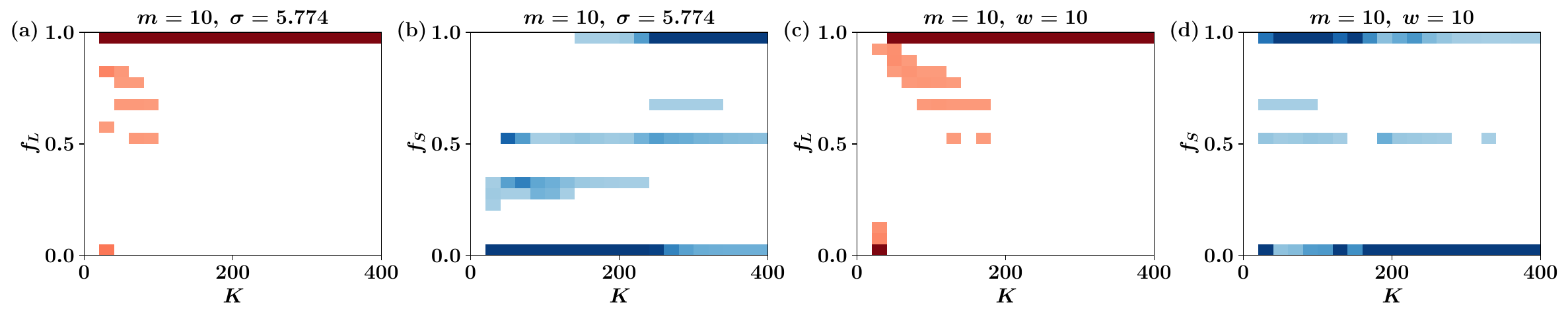}
\caption{
{\bf Seed-fraction statistics.}
Seed fractions $f_L$ and $f_S$ of large and small clusters—the proportion of final clusters that already exist as seed clusters at the seeding time $t_s$ [as defined in Fig.~\ref{fig:rev6}]—are plotted versus coupling strength $K$ for $m=10$. 
Panels (a,b) show results for the Gaussian case with $\sigma=5.774$, and (c,d) for the uniform case with $w=10$, displaying $f_L$ and $f_S$ respectively.
The large clusters exhibit similar timing in both distributions—$f_L$ is high in both cases [(a,c)], indicating that they are mostly in place by the seed stage—whereas the small clusters behave differently: for the Gaussian, $f_S$ increases with $K$, while for the uniform distribution it decreases with $K$ [(b,d)].
}
\label{fig:rev13}
\end{figure*}

\begin{figure*}[!t]
\centering
\includegraphics[width=1.0\linewidth]{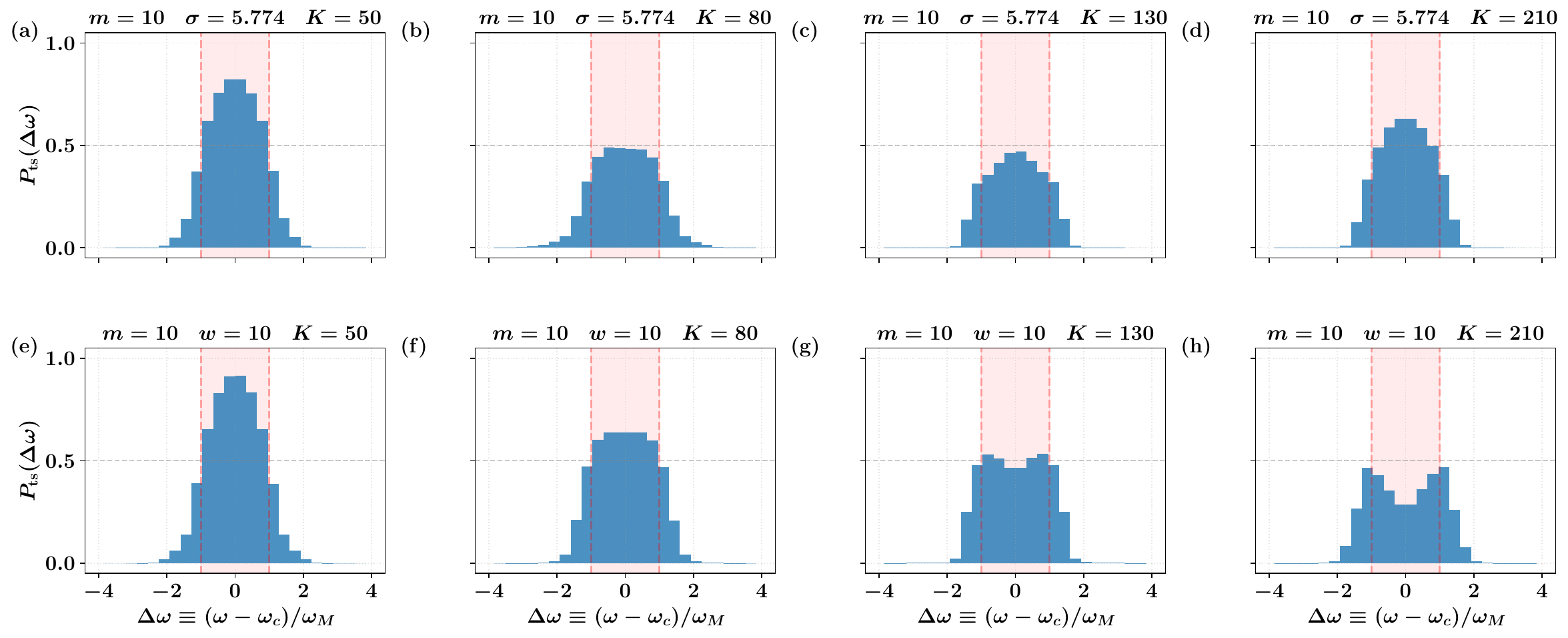}
\caption{
{\bf Transient capture probability and seed location.}
Transient capture probability $P_{\mathrm{ts}}(\Delta\omega)$ at time $t_s$ plotted against the normalized frequency detuning $\Delta\omega\equiv(\omega-\omega_c)/\omega_M$ for $m=10$; the shaded area denotes the Melnikov domain $|\Delta\omega|<1$. Here, $P_{\mathrm{ts}}$ represents the fraction of oscillators with offset $\Delta\omega$ that are already attached to a seed cluster, so its maximum identifies the frequency region where long-lived clusters originate.
(a--d) Gaussian $\sigma=5.774$ and (e--h) uniform $w=10$, at $K=50$, $80$, $130$, and $210$.
For the Gaussian case, $P_{\mathrm{ts}}$ remains unimodal with a maximum at $\Delta\omega = 0$ for all $K$ (central seeding). In contrast, for the uniform case, the increasingly strong merging effect drives the seeds away from the center as $K$ grows, causing $P_{\mathrm{ts}}$ to develop a dip at the origin and to split into two lobes around and beyond $|\Delta\omega| = 1$ (peripheral seeding).
}
\label{fig:rev9}
\end{figure*}

\begin{figure}[!t]
\centering
\includegraphics[width=1.0\linewidth]{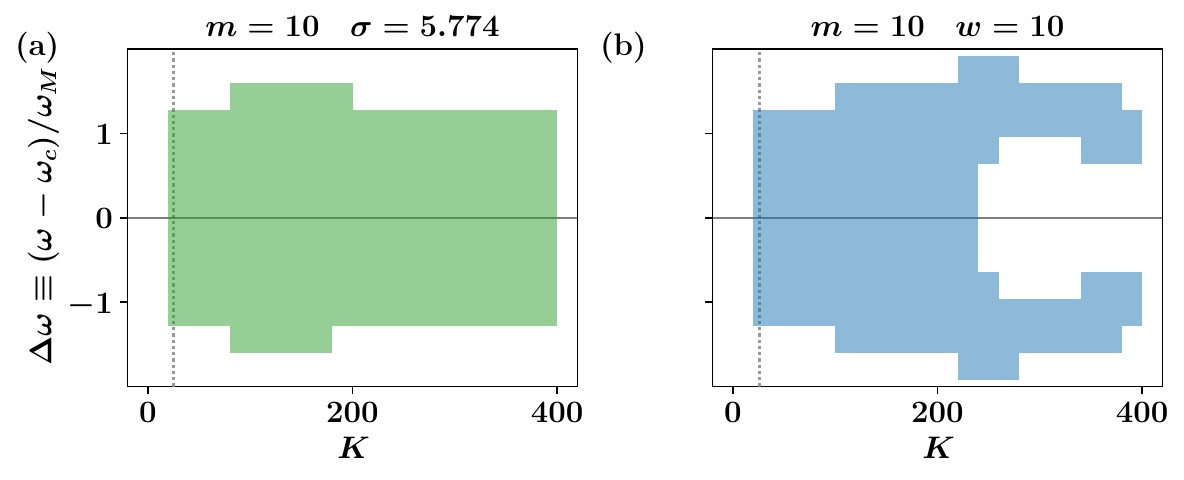}
\caption{
{\bf Migration of the seeding site with coupling.}
High-$P_{\mathrm{ts}}$ region in the normalized frequency offset $\Delta\omega\equiv(\omega-\omega_c)/\omega_M$ versus $K$, for $m=10$: (a) Gaussian $\sigma=5.774$ and (b) uniform $w=10$.
At each $K$, shaded bars mark the $\Delta\omega$ bins where the transient capture probability $P_{\mathrm{ts}}$ of the large clusters exceeds its mean; the grey dotted line is the critical coupling and the horizontal line is $\Delta\omega=0$.
For the Gaussian the high-$P_{\mathrm{ts}}$ region stays within the Melnikov region ($|\Delta\omega|\lesssim1$) at all $K$, whereas for the uniform it migrates beyond it as $K$ increases, confirming the shift from central to peripheral seeding.
}
\label{fig:rev10}
\end{figure}

\subsection*{Cluster Kinetics on the Microscopic Scale}

The contrasting hierarchical and homogeneous structures of the two $g(\omega)$ distributions motivate an examination of how their clusters evolve. Up to intermediate sizes, clusters in both cases grow in much the same way—a central core progressively draws in surrounding oscillators one at a time. They differ only in how this growth ultimately concludes. In the Gaussian case, the pronounced central peak generates a single, dominant cluster that grows larger than all others and progressively pulls the remaining oscillators into synchronization until it absorbs them, and $R$ rises smoothly. In contrast, for the uniform case, no single cluster dominates; instead, clusters of comparable size distributed across the frequency spectrum successively merge to form a giant. Each jump in the order parameter corresponds to a merger, during which $\Delta R$ exhibits sharp peaks because the large number of possible cluster configurations makes the final state highly sensitive to the initial conditions. Consequently, the behavior of $R$ and $\Delta R$ in Fig.~\ref{fig:rev1} already reflects the formation pathway, which we now proceed to reconstruct explicitly.

Figure~\ref{fig:rev4} depicts the temporal evolution of $\langle\dot\theta_i\rangle_\tau$ to show merging kinetics for the two $g(\omega)$ distributions. For the Gaussian case in Figure~\ref{fig:rev4}(a-d), a single large cluster first forms at the center and then grows outward by entraining neighboring oscillators, with smaller clusters following only afterward; for the uniform case, several clusters of comparable size nucleate nearly simultaneously and subsequently coalesce through merging [Fig.~\ref{fig:rev4}(e--h)]. Merger becomes more important as the coupling increases: For larger values of $K$, interactions between clusters become dominant, and the giant cluster forms primarily through the coalescence of existing seed clusters rather than through the entrainment of individual oscillators. As a result, the seeds of this giant cluster are located progressively farther from the center, extending toward the periphery.

For the following analysis of cluster evolution at a given $K$ and fixed initial conditions, we introduce three characteristic times $t_i$, $t_s$, and $t_f$ [Fig.~\ref{fig:rev6}]. At $t_i$, no cluster has yet formed; $t_f$ is the final time defined in Materials and Methods, at which the cluster profiles have become stationary. We term the clusters identified at $t_s$ and $t_f$ seed clusters and clusters, respectively. A cluster at $t_f$ is large if its size is at least half that of the largest cluster, $s_{\max}/2$, and small otherwise; $t_s$ is then the earliest time at which the number of large seed clusters exceeds that of large clusters. A seed cluster at $t_s$ is large if its size is at least $2\sqrt{N}$, and small otherwise; this threshold accommodates the size fluctuations of cluster aggregation and fragmentation~\cite{kang1984scaling} and ensures that a split seed cluster still maps onto a large cluster [Fig.~\ref{fig:rev6}].

To characterize when the final clusters are established, we measure the seeded fractions $f_L$ and $f_S$, the proportions of large and small clusters at $t_f$ that map onto a seed cluster at $t_s$; two clusters map to each other when they share at least $\sqrt{N}$ oscillators [Fig.~\ref{fig:rev13}]. The large clusters emerge concurrently for both shapes, with $f_L$ high [Fig.~\ref{fig:rev13}(a,c)], so they are already established by $t_s$. The small clusters instead behave oppositely: $f_S$ approaches unity at large $K$ for the Gaussian but at small $K$ for the uniform [Fig.~\ref{fig:rev13}(b,d)]. The cluster kinetics behind this reversal are described below.

In the Gaussian case, when $K$ is small, the oscillators in the periphery are unable to be members of the single dominant cluster in the dense core. But they form clusters independently. Later, they can become members after the dominant cluster has already developed and decelerated them near its far side, causing them to miss the initial nucleation stage. When $K$ is large, such oscillators instead organize into their own independent small cluster and they already appear during the nucleation stage, causing $f_S$ to increase.

In the uniform case, clusters of every size appear simultaneously over the entire frequency range when $K$ is small, so small clusters are already present during nucleation. When $K$ becomes large, however, these clusters coalesce into a single dominant cluster. Only the remaining oscillators subsequently assemble into small clusters after the dominant cluster has formed and decelerated them near its far side. As a result, they bypass the nucleation stage, and $f_S$ is reduced.

In both cases, the small clusters are never seen before the large ones; instead, they appear either concurrently with the large clusters or only after them. Their nucleation can therefore proceed in sequence rather than all at once. This ordering in both size and formation time persists as inertia and variance are changed, providing a robust signature of the specific shape of the distribution. [Sec.~\ref{sec:RS5f}].

We examine how the large seed clusters identified at $t_s$ evolve [Fig.~\ref{fig:rev6}]. To trace each large seed cluster to the clusters it ultimately joins, we connect it to any cluster sharing at least $\sqrt{N}$ oscillators with it [Fig.~\ref{fig:rev6}(b)]. A large seed cluster usually maps one-to-one onto a single cluster, but it may occasionally \emph{split} into two, or two seed clusters may \emph{merge} into one, so the mapping is generally many-to-many. These splitting and merging events rearrange the seed clusters into their final configuration and distinguish the two routes: the mapping stays one-to-one under entrainment but becomes many-to-one under merger.

We then ask where in frequency space each cluster was seeded. For a cluster $C$ with mean frequency $\omega_c$ and order parameter $R_c$, we collect the oscillators of all large seed clusters $S_a$ mapped to it,
\begin{equation}
\mathcal{S}(C) \;=\; \bigcup_{a\,:\,S_a\ \mathrm{large},\ |S_a\cap C|\,\geq\,\sqrt{N}} S_a .
\end{equation}
Each oscillator $i\in\mathcal{S}(C)$ enters at a normalized frequency offset $\Delta\omega_i\equiv(\omega_i-\omega_c)/\omega_M$, where $\omega_M(R_c,m,K)$ is the Melnikov half-width of $C$. Collecting these offsets over all clusters and realizations, we define the transient capture probability
\begin{equation}
P_{\mathrm{ts}}(\Delta\omega) \;=\; \frac{N_{\mathrm{obs}}(\Delta\omega)}{N_{\mathrm{exp}}(\Delta\omega)},
\end{equation}
where $N_{\mathrm{obs}}(\Delta\omega)$ counts the seed-bound oscillators in each $\Delta\omega$ bin and $N_{\mathrm{exp}}(\Delta\omega)=N\!\int_{\mathrm{bin}} g(\omega)\,d\omega$ is the number expected from the frequency distribution $g(\omega)$, with the bin mapped back through $\omega=\omega_c+\omega_M\,\Delta\omega$. Its peak marks where seeding most likely occurs. For the Gaussian, central assembly keeps $P_{\mathrm{ts}}$ unimodal and peaked at $\Delta\omega=0$ for all $K$; for the uniform, the seeds are pushed outward as $K$ grows, so $P_{\mathrm{ts}}$ develops a central dip and splits into two lobes near and beyond the Melnikov boundary $|\Delta\omega|=1$ [Fig.~\ref{fig:rev9}]. Condensing the full $K$-dependence, Fig.~\ref{fig:rev10} tracks the band of highest $P_{\mathrm{ts}}$: for the Gaussian it stays within the Melnikov region ($|\Delta\omega|\lesssim 1$), whereas for the uniform it migrates beyond it. This central-versus-peripheral seeding holds across inertia and variance, a robust signature of the distribution shape [Sec.~\ref{sec:RS8}].

\begin{figure}[!t]
\centering
\includegraphics[width=1.0\linewidth]{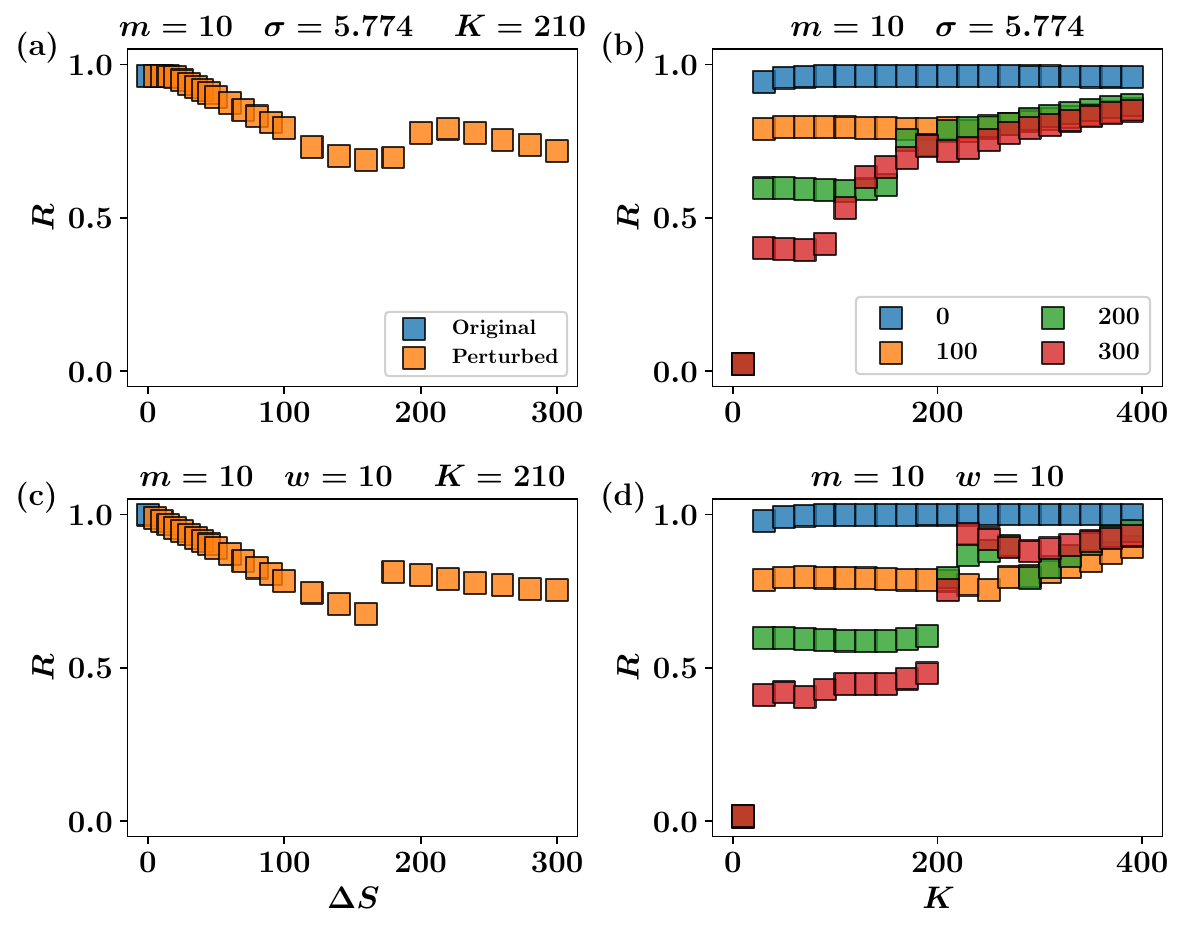}
\caption{
{\bf Robustness to peripheral perturbations.}
The system is prepared in a maximally synchronized state by a backward process; the $s$ lowest and $s$ highest intrinsic-frequency oscillators are then perturbed by resetting their velocities to $\dot\theta_i=\omega_i/\gamma$ and randomizing their phases.
(a,c) Order parameter $R$ versus perturbation size $s$ at fixed $K=210$, comparing original and perturbed states, for (a) Gaussian with $\sigma=5.774$ and (c) uniform with $w=10$.
(b,d) $R$ versus $K$ for perturbation sizes $s=0$, $100$, $200$, and $300$, for (b) Gaussian and (d) uniform with the same variance, respectively.
The two shapes are robust in opposite coupling ranges: the Gaussian recovers readily in the range from weak to intermediate $K$, where the uniform stays trapped in a low-$R$ plateau, whereas at strong $K$ the uniform also recovers and the perturbed--unperturbed gap closes.
}
\label{fig:rev11}
\end{figure}

\subsection*{Robustness or Vulnerability of the Largest Cluster}

The aggregation pathway likewise determines how robust the synchronized state is against perturbations. To create a maximally synchronized configuration, we employ a backward protocol: a fully locked state prepared at large coupling is used as the initial condition at the target $K$ [Materials and Methods], and the system is then perturbed by picking the $s$ oscillators with the smallest intrinsic frequencies and the $s$ oscillators with the largest intrinsic frequencies, resetting their velocities to $\dot\theta_i=\omega_i/\gamma$ and randomizing their phases [Fig.~\ref{fig:rev11}].

For the Gaussian case, the giant cluster recovers the peripheral oscillators through individual re-entrainment. On the other hand, for the uniform case, it recovers the peripheral oscillators through cluster merging.

For weak or intermediate $K$, it is easier to pull in individual oscillators, and the cluster is readily reformed in the Gaussian case. In contrast, the uniform distribution remains in the plateau region with low $R$ values, where the reformed clusters merge slowly, leaving a wide gap from the unperturbed curve [Fig.~\ref{fig:rev11}(b,d)]. For strong $K$, this balance is reversed. Now the entire clusters can recombine more easily, so the uniform distribution is restored and the gap disappears.

Stability is thus inherited from the aggregation pathway: entrainment, working oscillator by oscillator, dominates at weak coupling, whereas merger, working cluster by cluster, dominates at strong coupling. Since the pathway is fixed by the shape at constant variance, it is ultimately the shape that governs stability---the final link in the chain running from distribution shape through cluster organization and assembly pathway to system robustness. This opposite-range robustness holds across inertia and variance, a robust signature of the distribution shape [Sec.~\ref{sec:RS9}].

\section*{Summary and Discussion}

We have studied the kinetics of synchronized clusters and the mechanism by which oscillators reach macroscopic synchronization in the 2nd KM. In the inertia-free 1st KM, the shape of the natural-frequency distribution plays only a limited role: it fixes the character of the transition---continuous for a Gaussian, mixed-order for a uniform distribution---but once past the transition point, both distributions relax in the same way, with a single macroscopic cluster growing smoothly toward full synchronization. Inertia changes this. By allowing multiple clusters with distinct rotational frequencies to coexist and interact, it turns synchronization into a multi-frequency-synchronized cluster, interaction-driven process, raising the question of whether the distribution still acts only on the transition or now reshapes the entire collective dynamics. We find the latter: the shape of the distribution---the presence or absence of a dominant central peak---governs the dynamics throughout, fixing the cluster organization, the assembly pathway, and ultimately the system's robustness.

Inertia first breaks the conventional mean-field description beyond the transition point: the order parameter grows not smoothly but through discrete jumps, its susceptibility $\Delta R$ spiking at each one as nearly identical initial conditions diverge into different final states. The shape's role must therefore be understood at the mesoscopic level of the clusters, where it sets both structure and route. A Gaussian, with its dense central peak, organizes hierarchically around one dominant cluster with unequal gap ratios such as $2\!:\!1$ and grows that cluster by entraining oscillators outward from the center. A uniform distribution, lacking a center, forms clusters of comparable size with equal $1\!:\!1$ gaps---a homogeneous DS---and assembles its giant cluster by merging peripheral seeds. This pathway in turn fixes the stability: the Gaussian recovers oscillator by oscillator through re-entrainment and is thus robust from weak to intermediate $K$, whereas the uniform recovers cluster by cluster through merger and is thus robust at strong $K$. A single structural feature---the central peak---thus runs through the entire chain, from cluster organization through the assembly pathway to system robustness [Table~\ref{tab:comparison}].

Two distinct synchronization pathways have also been reported in the 1st KM~\cite{gomez2007paths}, albeit in a different setting: in random networks, small locally synchronized seed clusters form and then merge, whereas in scale-free networks, a synchronized central core emerges first and progressively entrains the surrounding oscillators. These resemble the merger- and entrainment-driven pathways found here, with one essential difference---there the route is set by the network structure, whereas here it is set by the shape of the frequency distribution. Moreover, those processes are overdamped, so their microscopic dynamics are comparatively simple next to the underdamped, multi-cluster dynamics of the 2nd KM.

These findings have potential implications for modern power grids. As intermittent renewable sources replace conventional generators, the effective frequency distribution broadens from Gaussian-like toward more uniform profiles---a change in both shape and variance of the kind studied here. Our analysis suggests that such grids increasingly rely on merger-driven synchronization, whose robustness rests on whole groups of high-power peripheral buses---typically large-scale renewable installations or critical interconnections---rejoining together rather than on individual nodes re-entraining. Where the effective coupling is not strong enough to draw these groups back, disruptions to such nodes can desynchronize the grid more readily than in conventional systems, so grid operators may need to prioritize protection and redundancy for these peripheral buses.
\begin{table*}[!t]
\centering
\footnotesize
\setlength{\tabcolsep}{4pt}
\renewcommand{\arraystretch}{1.3}
\begin{tabular}{p{3.4cm}p{4.9cm}p{4.9cm}}
\hline
\hline
 & \textbf{Gaussian} & \textbf{Uniform} \\
\hline
\multicolumn{3}{l}{\textit{Macroscopic Dynamics}} \\
Order parameter & Smooth growth, single susceptibility peak & Stepwise growth, peak at every jump \\
Self-consistency & Locates $K_c$; fails beyond transition point & Locates $K_c$; fails beyond transition point \\
Multistability & \multicolumn{2}{c}{Wide window for both; width set by variance and inertia} \\
\hline
\multicolumn{3}{l}{\textit{Cluster Organization}} \\
Structure & Centralized & Decentralized \\
Cluster spacing & Hierarchical ratios (e.g.\ $2\!:\!1$) & Equal ratios ($1\!:\!1$) \\
\hline
\multicolumn{3}{l}{\textit{Microscopic Mechanism}} \\
Binding determinant & Initial energy & Initial energy \\
Seed location & Central & Central to peripheral \\
Pathway & Entrainment & Entrainment to merger \\
\hline
\multicolumn{3}{l}{\textit{System Stability (under peripheral perturbation)}} \\
Recovery unit & Single oscillators & Whole clusters \\
Robust range & Weak to intermediate $K$ & Strong $K$ \\
\hline
\multicolumn{3}{l}{\textit{Physical Systems}} \\
Representative & Conventional grids & Renewable-integrated grids \\
\hline
\hline
\end{tabular}
\caption{{\bf Gaussian versus uniform: key distinctions.} Summary of synchronization characteristics in the second-order Kuramoto model with inertia, comparing macroscopic dynamics, cluster organization, microscopic mechanisms, and system stability.}
\label{tab:comparison}
\end{table*}

\section*{Materials and Methods}

We simulate the 2nd KM on a fully connected network of $N = 1024$ oscillators, with damping fixed at $\gamma = 1$ and the inertia $m$ and coupling $K$ specified in the text. Each intrinsic frequency $\omega_i$ is drawn from a zero-mean distribution $g(\omega)$, uniform or Gaussian, whose width (the uniform half-width $w$ or the Gaussian standard deviation $\sigma$) is specified in the text; the two are compared at matched variance $\sigma = w/\sqrt{3}$. In the forward process, used for the synchronization transition and the formation analysis, each phase $\theta_i(0)$ is sampled uniformly on $[-\pi, \pi]$ and the initial angular velocity is set to $\dot{\theta}_i(0) = \omega_i/\gamma$, following previous studies~\cite{tanaka1997first, tanaka1997self}. In the backward process, used to prepare the maximally synchronized state for the robustness analysis, each realization is first driven to full synchronization at a large coupling $K=800$: starting from $\theta_i(0)$ uniform on $[-\pi,\pi]$ and $\dot\theta_i(0)=\omega_i/\gamma$, it is integrated for $3200$ time units until a fully locked state is reached. The coupling is then set directly to the target value $K$, and the fully locked configuration is used as the initial condition; the system is integrated for a further $6400$ time units, and steady-state quantities are averaged over the final interval $t\in[5600,6400]$. The equations of motion are integrated by the fourth-order Runge-Kutta method with a fixed time step $\Delta t = 0.01$. Unless stated otherwise, each realization is integrated up to the final time $t_f=10400$, and all steady-state quantities---the order parameter $R$, the time-averaged angular velocities $\langle\dot\theta_i\rangle$, the cluster number fractions $\rho_L$ and $\rho_S$, and the Devil's-staircase statistics---are obtained by time-averaging over the final interval $t\in[9600,t_f]$, by which time the cluster profiles no longer change. Synchronized clusters are identified from the steady-state angular velocities $\langle\dot\theta_i\rangle$ by grouping oscillators with matching velocities and merging nearby groups; the procedure and its tolerances are detailed in Sec.~\ref{sec:cluster_id}.

During manuscript preparation, we used a large language model to improve the clarity and readability of the text. All AI-assisted content was carefully reviewed and validated by the authors, who maintain full responsibility for the scientific content, accuracy, and integrity of this work.


\section*{Acknowledgments:}
We gratefully acknowledge the anonymous reviewer for the valuable suggestion to employ equal standard deviations.
\paragraph*{Funding:}
B.K. was supported by the National Research Foundation of Korea by Grant No. RS-2023-00279802 and the KENTECH Research Grant No. KRG-2021-01-007.
S.B. acknowledges support from the project n.PGR01177 of the Italian Ministry of Foreign Affairs and International Cooperation.
\paragraph*{Author contributions:}
Conceptualization: C.H.K., Methodology, Investigation: C.H.K. and Y.K., Visualization: C.H.K. and Y.K., Supervision: B.K., Writing--original draft: C.H.K., Writing--review and editing: B.K. and S.B.. 
\paragraph*{Competing interests:}
There are no competing interests to declare.
\paragraph*{Data and materials availability:}
All data needed to evaluate the conclusions in the paper are present in the paper and/or the Supplementary Materials. Additional data related to this paper may be requested from the authors.


\putbib
\end{bibunit}

\clearpage
\newpage

\appendix

\onecolumngrid


\makeatletter
\renewcommand{\thesection}{S\arabic{section}}
\renewcommand{\theequation}{S\arabic{equation}}
\renewcommand{\thefigure}{S\arabic{figure}}
\renewcommand{\thetable}{S\arabic{table}}
\renewcommand{\bibnumfmt}[1]{[S#1]}

\setcounter{section}{0}
\setcounter{equation}{0}
\setcounter{figure}{0}
\setcounter{table}{0}
\setcounter{page}{1}
\makeatother

\begin{center}
\textbf{\large Supplementary Material for \\ Paths to synchronization in the Kuramoto model with inertia}
\end{center}

\begin{bibunit}



\section{Melnikov Derivation of the Entrainment Boundary}
\label{sec:melnikov_derivation}

Here we derive the leading term of the entrainment boundary $\omega_M$ that enters the self-consistency equation, Eq.~\eqref{eq:self_consistency} of the main text. The leading term $(4\gamma/\pi)\sqrt{KR/m}$ follows from the Melnikov method, as shown below; the subleading term $\alpha\sqrt{1/(KRm^{3})}$ is not derived analytically but is an empirical correction fitted to the numerics, whose origin we clarify at the end.

Consider an oscillator $i$ rotating with a finite angular velocity $\dot\theta_i\neq0$. Its rotation can be maintained only if the total non-conservative energy $E_{\rm nc}$ remains non-negative,
\begin{align}
E_{\rm nc} \equiv W_{\omega} + W_{\gamma} \geq 0,
\end{align}
where $W_\omega$ is the energy supplied by the intrinsic frequency and $W_\gamma$ the energy dissipated by damping. The difficulty in applying this criterion is that $E_{\rm nc}$ depends on the full nonlinear trajectory, which cannot be computed exactly beyond a finite time, so the path-dependent quantities $W_\omega$ and $W_\gamma$ are inaccessible in general.

The Melnikov method addresses this by working in the limit $E_{\rm nc}\to0$, where the trajectory stays close to the unperturbed conservative orbit and the analysis becomes tractable. Let $(\theta(t),\dot\theta(t))$ denote the configuration at $E_{\rm nc}=0$. When $\omega$ and $\gamma$ are switched on, the trajectory shifts only slightly, to $(\theta_{\omega,\gamma},\dot\theta_{\omega,\gamma})$ with $\theta_{\omega,\gamma}\approx\theta$ and $\dot\theta_{\omega,\gamma}\approx\dot\theta$. On the unperturbed orbit the angular velocity obeys
\begin{align}
\dot{\theta} = \sqrt{\frac{2KR}{m}\left(1 - \cos\theta\right)}
            = 2\sqrt{\frac{KR}{m}}\,\sech\!\left(\sqrt{\frac{KR}{m}}\,t\right),
\label{eq:trajectory_0}
\end{align}
the separatrix solution of the conservative pendulum. Over one revolution, the net non-conservative energy is
\begin{align}
E_{\rm nc}
&= \int_{0}^{2\pi} d\theta\,\omega \;-\; \int_{0}^{2\pi} d\theta\,\gamma\dot\theta \cr
&= 2\pi\omega \;-\; \gamma\int_{-\infty}^{\infty} dt\,\dot\theta^{2} \cr
&= 2\pi\omega \;-\; 4\gamma\,\frac{KR}{m}\int_{-\infty}^{\infty} dt\,\sech^{2}\!\left(\sqrt{\frac{KR}{m}}\,t\right) \cr
&= 2\pi\omega \;-\; 8\gamma\sqrt{\frac{KR}{m}},
\end{align}
where we used $d\theta=\dot\theta\,dt$, so that $\int d\theta\,\gamma\dot\theta=\gamma\int dt\,\dot\theta^{2}$, and then substituted Eq.~\eqref{eq:trajectory_0} together with $\int_{-\infty}^{\infty}\sech^{2}(bt)\,dt=2/b$. The oscillator can no longer sustain its rotation once $E_{\rm nc}<0$, that is, once
\begin{align}
\omega \le \frac{4\gamma}{\pi}\sqrt{\frac{KR}{m}}.
\end{align}
For a cluster with mean frequency $\omega_c$ and order parameter $R$, the same argument gives the entrainment condition
\begin{align}
|\omega-\omega_c| \le \frac{4\gamma}{\pi}\sqrt{\frac{KR}{m}},
\end{align}
which is the leading term of $\omega_M$.

This leading-order boundary captures the essential physics but deviates slightly from the numerics. To account for this deviation, we include an empirical correction obtained by fitting to numerical data~\cite{belykh2016bistability, strogatz2018nonlinear, gao2018self}, giving the full boundary used in the main text,
\begin{align}
\omega_M = \frac{4\gamma}{\pi}\sqrt{\frac{KR}{m}} + \alpha\sqrt{\frac{1}{KRm^{3}}},
\qquad \alpha \approx -0.3056 .
\end{align}
The first term is the Melnikov result derived above; the second is the empirical correction, with $\alpha \approx -0.3056$ fixed numerically. This boundary is a leading-order approximation, derived for a single isolated cluster on the near-separatrix orbit and valid only in that regime; it serves as a reference boundary rather than an exact criterion. As shown in Sec.~\ref{sec:melnikov_failure}, once several comparable clusters coexist and interact---the regime in which most of our analysis lies---this single-cluster criterion no longer determines cluster membership, which is instead set by the initial energy [Sec.~\ref{sec:initial_energy}].

\clearpage
\newpage


\section{Self-Consistency Equation and the ad-hoc Potential}
\label{sec:adhoc}

\begin{figure}[h]
\centering
\includegraphics[width=1.0\linewidth]{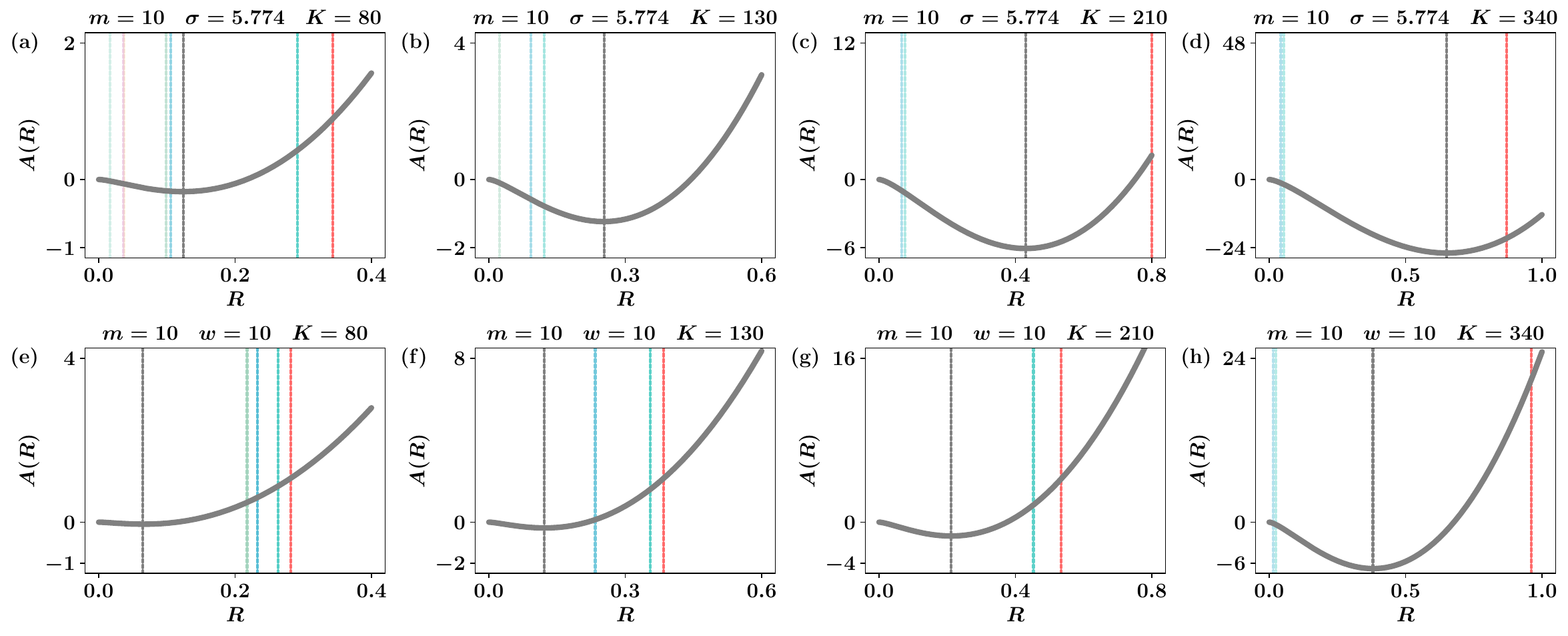}
\caption{
{\bf Breakdown of the ad-hoc potential model beyond the transition point.}
Ad-hoc potential $A(R)$ obtained under the assumption that the cluster grows by entraining oscillators one by one outward from the center of the frequency distribution (grey curve), with its theoretical minimum $R^*$ (grey dotted line), compared with the realized cluster order parameters $R_c$ (colored vertical lines, one per cluster; opacity scales with cluster size) for $m=10$.
(a--d) Gaussian $\sigma=5.774$ and (e--h) uniform $w=10$, at $K=80$, $130$, $210$, and $340$.
For both distributions the order parameter is carried not by a single growing cluster but by several coexisting clusters (multiple colored lines)---organized hierarchically around a dominant central cluster for the Gaussian and into clusters of comparable size for the uniform---and in neither case does the dominant cluster's $R_c$ coincide with the predicted minimum $R^*$ at any coupling shown.
}
\label{fig:figS2}
\end{figure}

In general, the self-consistency equation for the order parameter $R$ is
\begin{align}
    R = \int_{\omega_\min}^{\omega_\max} d\omega \, g(\omega) \sqrt{1 - \left(\frac{\omega - \omega_{c}}{KR}\right)^2}.
\end{align}
This equation assumes that the synchronized cluster grows by sequentially entraining oscillators with nearby intrinsic frequencies. The integration limits $\omega_\min$ and $\omega_\max$ span the frequency range of entrained oscillators and evolve as the cluster grows; their form depends on the direction of the process: $\omega_{\min,\max} = \pm KR$ in the backward process, and $\omega_{\min,\max} = \pm (4/\pi)\sqrt{KR/m}$ in the forward process~\cite{tanaka1997first, tanaka1997self}.
This equation locates the critical coupling $K_c$ in both the 1st KM and the 2nd KM, and predicts the full evolution of $R$ in the 1st KM and in the narrow, low-inertia regime of the 2nd KM. As we now show, however, it begins to fail beyond the transition point once the inertia or the variance grows, regardless of the distribution shape.

To generalize the approach, we employ a macroscopic potential---the ad-hoc potential~\cite{song2020effective, jhun2022quantum}---which models the growth of $R$ through this sequential entrainment:
\begin{align}
    A(R) = \int_{0}^{R} dR' \left[ R' - \int_{\omega_{\min}}^{\omega_{\max}} g(\omega) \sqrt{1 - \left( \frac{\omega - \omega_{c}}{K R'} \right)^2} d\omega \right],
    \label{eq:adhoc_potential_supp}
\end{align}
where $\omega_{c}$ is the mean natural frequency of the synchronized cluster, and the integration limits take the same forms as above, centered on $\omega_{c}$. The potential $A(R)$ provides a landscape in which cluster growth minimizes the macroscopic energy, the stable cluster size being the location $R^*$ of its minimum. More broadly, describing the macroscopic states of a system through such a single-variable energy function has been widely employed across collective systems~\cite{jang2015ashkin, kim2024entropy, kim2026heterogeneous}.

This picture holds only at the transition point. The self-consistency equation correctly predicts the coupling $K_c$ at which a cluster first nucleates (main text), but not the subsequent growth of $R$. Figure~\ref{fig:figS2} makes this explicit. Beyond the transition point, the order parameter is shared among several coexisting clusters rather than concentrated in one---organized hierarchically around a dominant central cluster for the Gaussian, into clusters of comparable size for the uniform---and in neither case does the dominant cluster's $R_c$ match the predicted minimum $R^*$ at any coupling shown. The discrepancy persists across $K$ rather than appearing only in a particular range.

The root of this failure is the multi-cluster configuration itself. Once the order parameter is shared among several clusters, the single-well potential---which presumes a single cluster growing by sequential entrainment---cannot describe it, whichever shape the clusters take. The variance and inertia then set how severe the failure becomes: the larger either one, the more clusters coexist and the greater the deviation from the single-cluster prediction. Both the self-consistency equation and the ad-hoc potential therefore fail to describe the growth of $R$ beyond the transition point in this regime, motivating the velocity-driven, multi-cluster description of the main text.

\clearpage
\newpage


\section{Finite Size Effects and Nucleation Barriers}

\begin{figure}[h]
\centering
\includegraphics[width=1.0\linewidth]{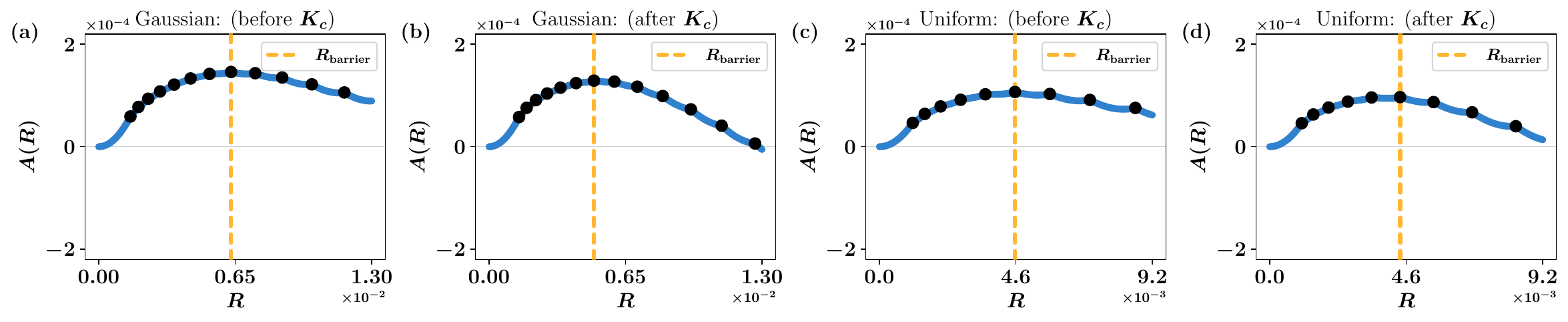}
\caption{
{\bf Finite-size nucleation barriers.}
Ad-hoc potential $A(R)$ near the transition point, for the Gaussian ($m=10$, $\sigma=5.774$) and uniform ($m=10$, $w=10$) distributions.
(a,b) Gaussian, below and above the critical coupling $K_c$; (c,d) uniform, below and above $K_c$.
The orange dashed line marks the barrier location $R_{\mathrm{barrier}}$, and the black dots are the discrete $R$ values accessible to the finite system ($N=1024$).
In both distributions a local maximum---the nucleation barrier---separates the disordered state ($R\approx0$) from the synchronized state at the global minimum, and the system must cross it through finite-size fluctuations to synchronize.
}
\label{fig:figRS1}
\end{figure}

Near $K_c$, the ad-hoc potential develops a local maximum---a nucleation barrier---between the disordered state at $R\approx0$ and the synchronized state at the global minimum [Fig.~\ref{fig:figRS1}]. Mean-field theory, taken in the infinite-size limit, predicts the global minimum as the stable state, but a finite system cannot reach it directly: it must first cross this barrier, which is common to both distributions.

The crossing is driven by finite-size fluctuations. In a finite system ($N=1024$), the discreteness of the oscillators produces such fluctuations, and a cluster nucleates only once they are large enough to carry the state over the barrier---escaping $R\approx0$ and falling toward the global minimum. The critical coupling $K_c$ therefore marks not where synchronization first becomes energetically favorable, but where the barrier falls to the scale of these fluctuations. Below $K_c$ [Figs.~\ref{fig:figRS1}(a,c)] the barrier exceeds them and the system stays trapped near $R\approx0$; above $K_c$ [Figs.~\ref{fig:figRS1}(b,d)] it has shrunk enough for them to carry the system across, triggering spontaneous synchronization. Beyond this point the cluster grows and the ad-hoc potential itself ceases to describe the dynamics [Sec.~\ref{sec:adhoc}], but the transition point is set precisely by this barrier crossing.

\clearpage
\newpage


\section{Cluster Identification}
\label{sec:cluster_id}

The steady-state mesoscopic quantities---the cluster number fractions $\rho_L$ and $\rho_S$, and the Devil's-staircase statistics ($N_s$, $f_{1:1}$, $\mathcal{V}_{\Delta\dot\theta}$, $f_{\mathrm{stair}}$)---rest on a single operational definition of a synchronized cluster, which we fix here.

For each oscillator we compute the steady-state angular velocity $\langle\dot\theta_i\rangle$, time-averaged over the final interval $t\in[9600,t_f]$ [Materials and Methods]. Clusters are then extracted in three stages. First, the oscillators are sorted by $\langle\dot\theta_i\rangle$, and adjacent oscillators are linked when their velocities differ by less than $\epsilon_1=10^{-4}$; each connected group of at least $16$ members is taken as an initial cluster, and smaller groups are set aside. Second, two initial clusters are merged when their mean angular velocities differ by less than $\epsilon_2=10^{-2}$, repeated until no further merging occurs, after which any oscillator not yet assigned is attached to the cluster whose mean angular velocity lies within $\epsilon_2$ of it. Third, only the merged clusters with at least $\sqrt{N}=32$ members are retained; oscillators in smaller groups, or outside $\epsilon_2$ of every cluster, remain unclustered and are counted as desynchronized. The size $s$ of a cluster is its number of member oscillators, and $s_{\max}$ denotes the size of the largest cluster in a given realization.

A cluster is labeled large if $s\geq s_{\max}/2$ and small otherwise. The number fractions $\rho_L$ and $\rho_S$ are the fractions of clusters that are large and small, counted over the identified clusters so that $\rho_L+\rho_S=1$; desynchronized oscillators do not form clusters and are excluded from this count. The angular-velocity gaps $\Delta\dot\theta_k$ underlying the Devil's-staircase statistics are the differences between the mean angular velocities of adjacent clusters obtained after the merging stage.

The seed clusters used in the formation analysis [Figs.~\ref{fig:rev6}, \ref{fig:rev13}, \ref{fig:rev9}, and \ref{fig:rev10}] are identified at the earlier time $t_s$, when the angular velocities have not yet settled and the sorted-gap criterion above is not yet sharp. We therefore identify seed clusters from the short-window angular velocity $\langle\dot\theta_i\rangle_\tau$ [Fig.~\ref{fig:rev4}] by a kernel-density method: a Gaussian kernel-density estimate of the $\langle\dot\theta_i\rangle_\tau$ distribution is formed [\texttt{gaussian\_kde} with \texttt{bw\_method}$=0.05$], and each local maximum of the estimated density with height at least $0.05$ defines a seed cluster. Oscillators lying within $\pm0.10$ of a peak are assigned to it; a seed cluster is retained only if it contains at least $\sqrt{N}=32$ oscillators, and the remaining oscillators are left unassigned. A seed cluster is termed large if it contains at least $2\sqrt{N}=64$ oscillators and small otherwise, the threshold chosen so that a large seed splitting into two still leaves each fragment near the $\sqrt{N}$ oscillators needed to map onto a large cluster.

\clearpage
\newpage


\section{Consistency of Shape-Dependent Growth Across Inertia and Variance}
\label{sec:RS2}

\begin{figure}[h]
\centering
\includegraphics[width=1.0\linewidth]{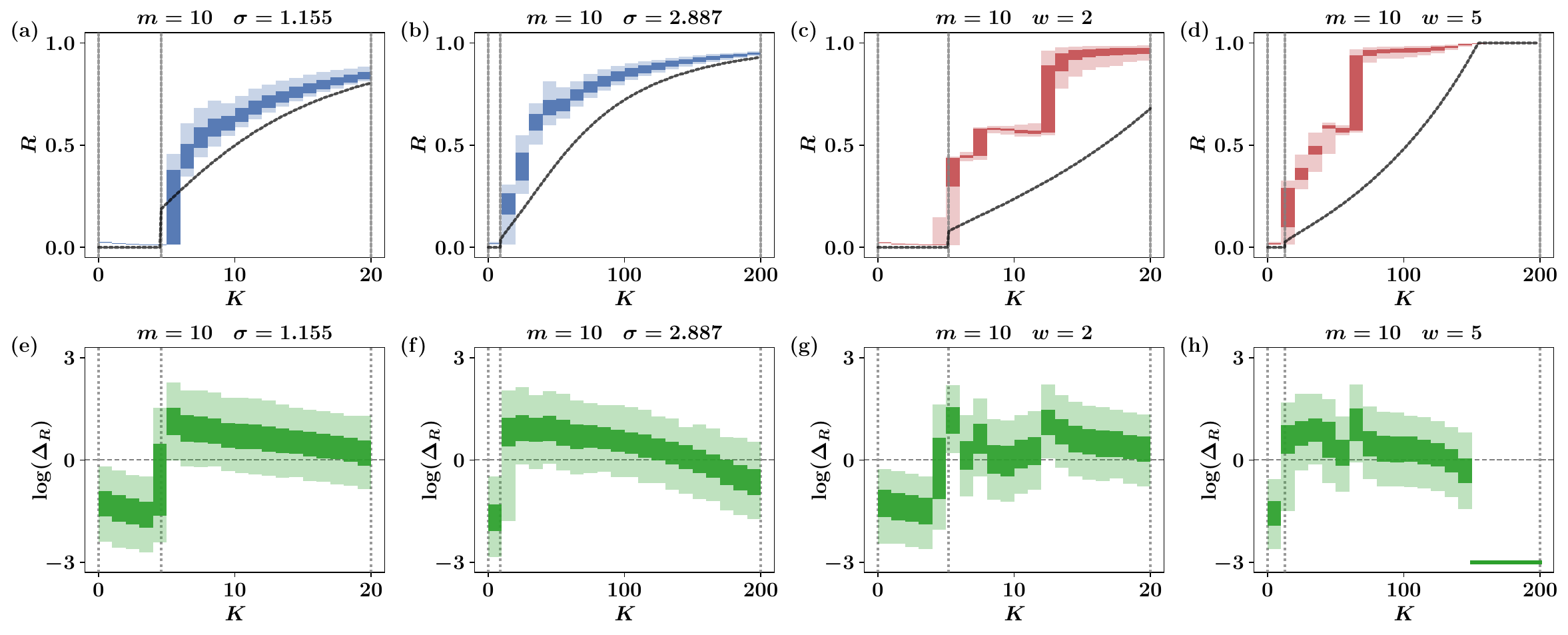}
\caption{
{\bf Shape-dependent growth across variance.}
Order parameter $R$ (a--d) and susceptibility $\log\Delta R$ (e--h) versus coupling strength $K$ for $m=10$, at varied variance: Gaussian $\sigma=1.155$, Gaussian $\sigma=2.887$, uniform $w=2$, and uniform $w=5$.
In (a--d) the black curve is the SCE solution of Eq.~\eqref{eq:self_consistency} and the shaded band is the interquartile range across realizations; in (e--h) the dark and light bands are the interquartile and $5$--$95\%$ ranges of $\log\Delta R$.
At every variance, the Gaussian rises smoothly with a single susceptibility peak at the transition point, whereas the uniform climbs in discrete jumps, each carrying its own peak.
}
\label{fig:figRS2a}
\end{figure}

\begin{figure}[h]
\centering
\includegraphics[width=1.0\linewidth]{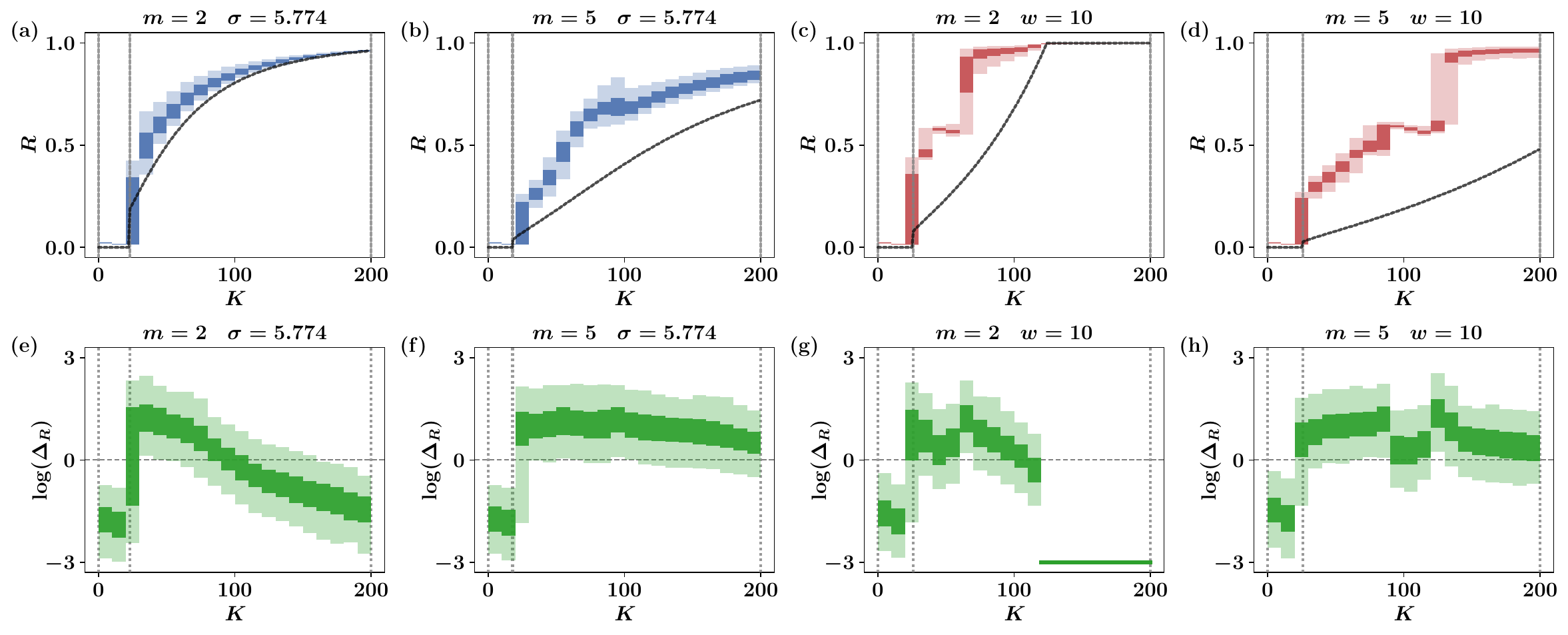}
\caption{
{\bf Shape-dependent growth across inertia.}
$R$ (a--d) and $\log\Delta R$ (e--h) versus $K$ at matched variance ($\sigma=5.774$, $w=10$), at varied inertia: Gaussian $m=2$, Gaussian $m=5$, uniform $m=2$, and uniform $m=5$.
Curves and bands follow Fig.~\ref{fig:figRS2a}.
The same contrast holds at every inertia: smooth, single-peak growth for the Gaussian and stepwise, multi-peak growth for the uniform.
}
\label{fig:figRS2b}
\end{figure}

The smooth-versus-stepwise contrast that separates the two shapes in the main text [Fig.~\ref{fig:rev1}] is not specific to the single inertia and variance shown there; it holds throughout parameter space. Figure~\ref{fig:figRS2a} fixes the inertia ($m=10$) and varies the variance, and Fig.~\ref{fig:figRS2b} fixes the variance and varies the inertia. In every case the Gaussian rises smoothly, its susceptibility $\log\Delta R$ peaking once at the transition point and decaying thereafter---the response of an ordinary transition with a single critical point---while the uniform grows through discrete jumps, each plateau marking a stable multi-cluster configuration and each jump a cluster merging event [Fig.~\ref{fig:rev5}], so that $\log\Delta R$ peaks anew at every jump as though it were a separate transition.

This contrast---one susceptibility peak for the Gaussian, one per jump for the uniform---is set by the presence or absence of a central peak in $g(\omega)$, and is therefore a robust signature of the distribution shape rather than of any particular inertia or variance.

\clearpage
\newpage


\section{Multistability and Initial-Condition Sensitivity}
\label{sec:multistability}

The main text shows that beyond the transition point the order parameter is initial-condition dependent: nearly identical initial states settle into measurably different final states, and the susceptibility $\Delta R$ rises above unity to register this sensitivity [Fig.~\ref{fig:rev1}]. This is the signature of a quenched multistability---the system becomes trapped in one of many coexisting states---and it arises for both the Gaussian and the uniform, differing only in the range of $K$ over which it persists. Here we characterize this multistability, define the susceptibility that measures it, and trace its dynamical origin.

\subsection{Multistability and the susceptibility \texorpdfstring{$\Delta R$}{Delta R}}

Two signatures in Fig.~\ref{fig:rev1} make this multistability concrete. The first is the spread of $R$ across realizations [shaded band, Fig.~\ref{fig:rev1}(a--d)]: within the multistable window, the interquartile range stays wide, the defining feature of coexisting stable fixed points into which nearly identical initial states settle apart. The second is the susceptibility behind this spread,
\begin{align}
\Delta R = \left\langle \frac{\lvert R^{\mathrm{final}} - \langle R^{\mathrm{final}}\rangle\rvert}{\lvert R^{\mathrm{initial}} - \langle R^{\mathrm{initial}}\rangle\rvert} \right\rangle,
\end{align}
the ensemble-averaged ratio of the final to the initial deviation of $R$ from its mean: $\log\Delta R \to -\infty$ means the system loses memory of its initial spread and converges, whereas $\log\Delta R > 0$ means infinitesimal initial differences are amplified, marking a coupling at which the system is critically sensitive [Fig.~\ref{fig:rev1}(e--h)].

The shape and the variance act on different facets of the dynamics, and on different parts of the $K$ axis. The shape fixes the character of the growth throughout: the uniform climbs in discrete jumps, each carrying its own susceptibility peak, while the Gaussian rises smoothly with a single peak at the transition point---a contrast that persists across inertia and variance [Sec.~\ref{sec:RS2}]. The variance instead sets the multistable window, the band of couplings over which $R$ stays initial-condition dependent: narrow near the transition point at small variance, it widens as the variance grows, for both shapes. The shape thus sets \emph{whether} the growth is stepwise or smooth, the variance only \emph{over what range of $K$} the multistability extends.

\subsection{Dynamical origin}

\begin{figure}[h]
\centering
\includegraphics[width=1.0\linewidth]{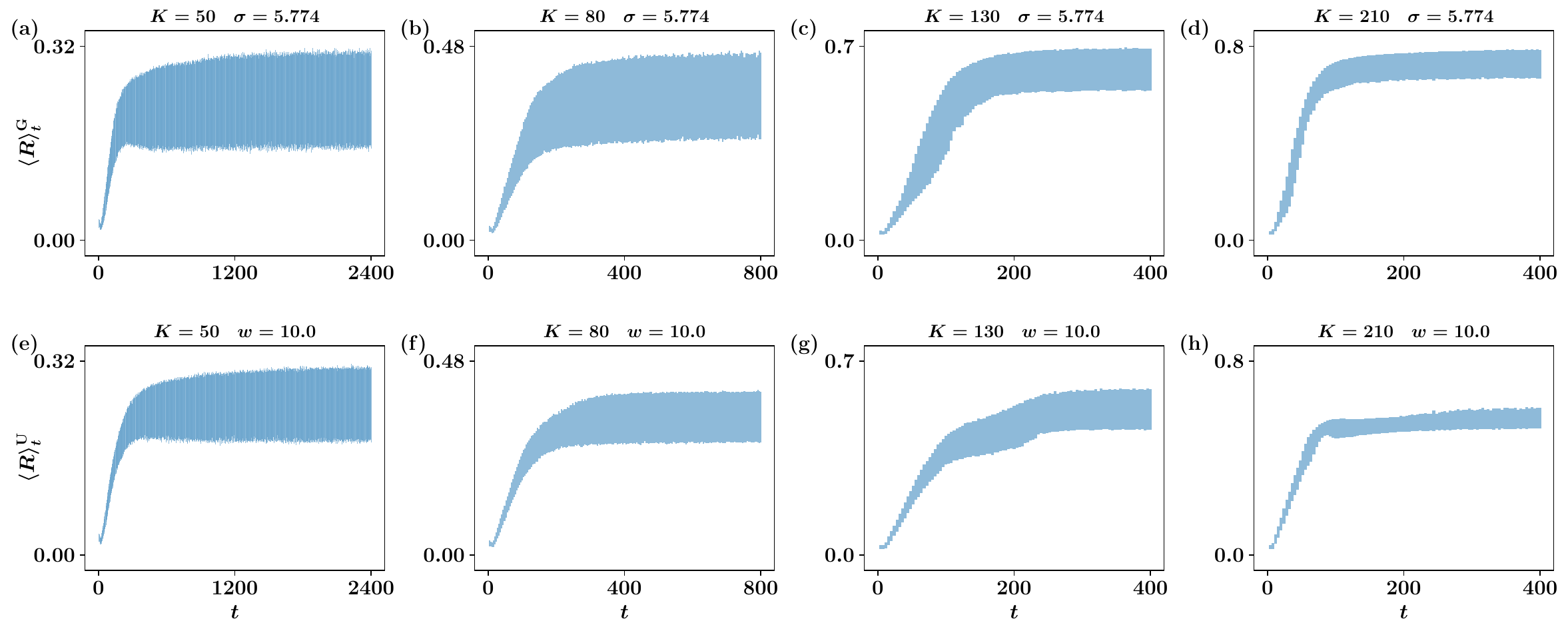}
\caption{
{\bf Temporal divergence of the order parameter across realizations.}
Spread of the time-averaged order parameter $\langle R\rangle_t$ versus time $t$, the $5$--$95\%$ band over $512$ realizations differing only in the random seed, for $m=10$.
(a--d) Gaussian $\sigma=5.774$ and (e--h) uniform $w=10$, at $K=50$, $80$, $130$, and $210$, all within the multistable window.
The band is narrow at early times and widens as the evolution proceeds, showing that small initial differences are amplified into distinct steady states.
Note the different time axes between columns.
}
\label{fig:revsmA}
\end{figure}

\begin{figure}[h]
\centering
\includegraphics[width=1.0\linewidth]{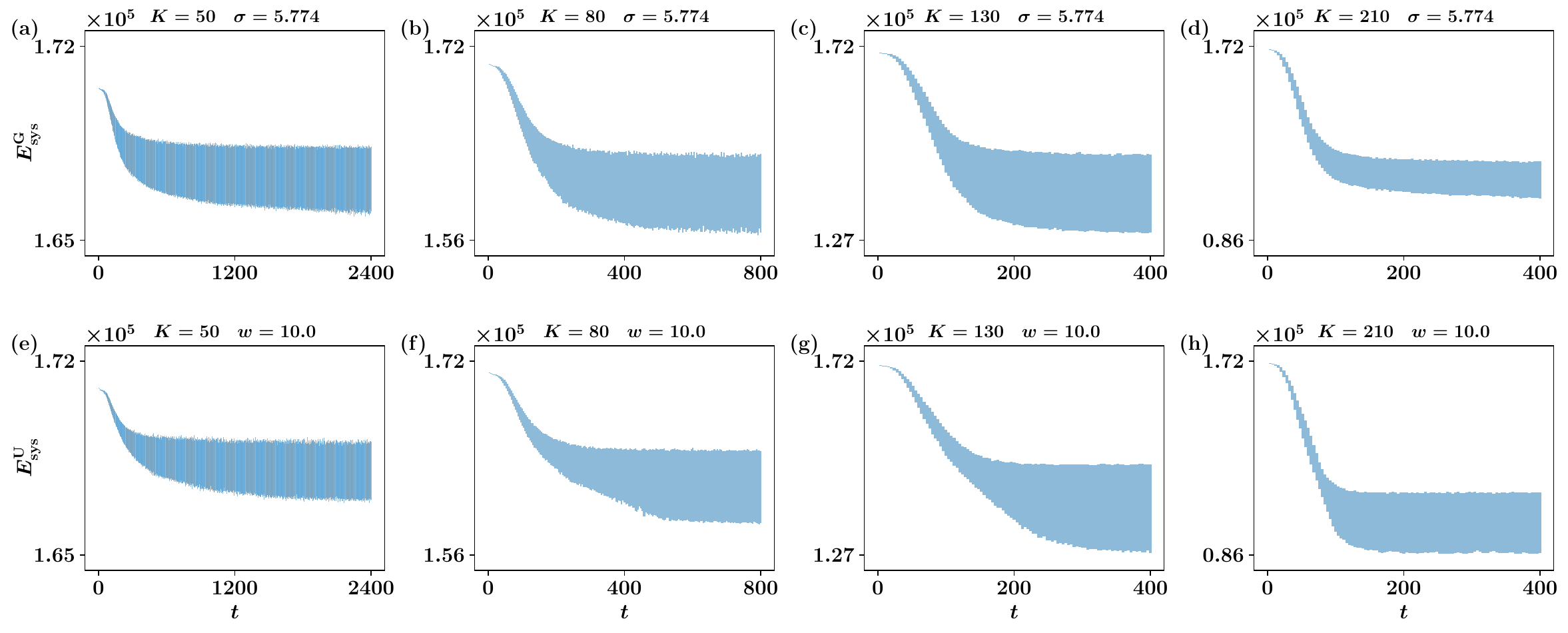}
\caption{
{\bf Temporal divergence of the total energy across realizations.}
Spread of the total energy $E_{\mathrm{sys}}$ versus time $t$, the $5$--$95\%$ band over $512$ realizations differing only in the random seed, for $m=10$.
(a--d) Gaussian $\sigma=5.774$ and (e--h) uniform $w=10$, at $K=50$, $80$, $130$, and $210$, all within the multistable window.
As in Fig.~\ref{fig:revsmA}, the band widens over time, confirming that the diverging realizations settle into genuinely different final states rather than rearranging at fixed macroscopic energy.
}
\label{fig:revsmB}
\end{figure}

We trace the origin of this sensitivity by following the time evolution of the ensemble rather than only its final state. Figures~\ref{fig:revsmA} and~\ref{fig:revsmB} show the spread across realizations of the time-averaged order parameter $\langle R\rangle_t$ and the total energy $E_{\mathrm{sys}}$, respectively, for the matched Gaussian ($\sigma=5.774$) and uniform ($w=10$) distributions at several couplings. The total energy is the sum of the kinetic and interaction contributions,
\begin{align}
    E_{\mathrm{sys}} = \frac{1}{2}m\sum_{i=1}^{N}\dot\theta_i^{2} - \frac{K}{2N}\sum_{i,j=1}^{N}\cos(\theta_j-\theta_i),
\label{eq:Esys}
\end{align}
where the first term is the rotational kinetic energy carried by the inertia and the second is the coupling energy. The driving term $\sum_i\omega_i\theta_i$ is excluded, as it grows without bound under the sustained rotation of the oscillators and would obscure the differences between final states. Both $E_{\mathrm{sys}}$ and the order parameter are shown after averaging over the short window $\tau$ defined in Fig.~\ref{fig:rev4}. Each shaded band is the $5$--$95\%$ interval over $512$ realizations that start from nearly identical initial conditions, differing only in the random seed.

At early times the bands are narrow, all realizations evolving together from statistically equivalent states. As time proceeds the bands widen, so that infinitesimal initial differences are amplified rather than damped---the dynamical origin of the susceptibility peaks of $\Delta R$. Trajectories that were initially indistinguishable diverge and settle into distinct steady states, each a different cluster configuration. Because this divergence appears in both the order parameter [Fig.~\ref{fig:revsmA}] and the total energy [Fig.~\ref{fig:revsmB}], the realizations reach genuinely different final states rather than rearranging at a fixed macroscopic value.

This asymptotic spread is the time-resolved counterpart of the multistability above: it is broad only within the multistable window, not at every $K$. The couplings shown ($K=50$ to $210$) all fall within that window for both shapes, and across this range the spread widens with the variance. Figures~\ref{fig:revsmA} and~\ref{fig:revsmB} thus resolve in time the sensitivity that the main text reports as $\Delta R$, the sensitivity that ultimately invalidates the single mean-field description.

\clearpage
\newpage


\section{Failure of the Melnikov Criterion}
\label{sec:melnikov_failure}

The SCE in Eq.~\eqref{eq:self_consistency} asserts that capture depends only on frequency proximity to $\omega_c$: within the Melnikov framework, oscillators with $|\omega-\omega_c|<\omega_M$ are captured with probability one and those outside are not, irrespective of the initial state. We test this with two diagnostics, both contrasting the matched Gaussian and uniform distributions.

\begin{figure}[h]
\centering
\includegraphics[width=1.0\linewidth]{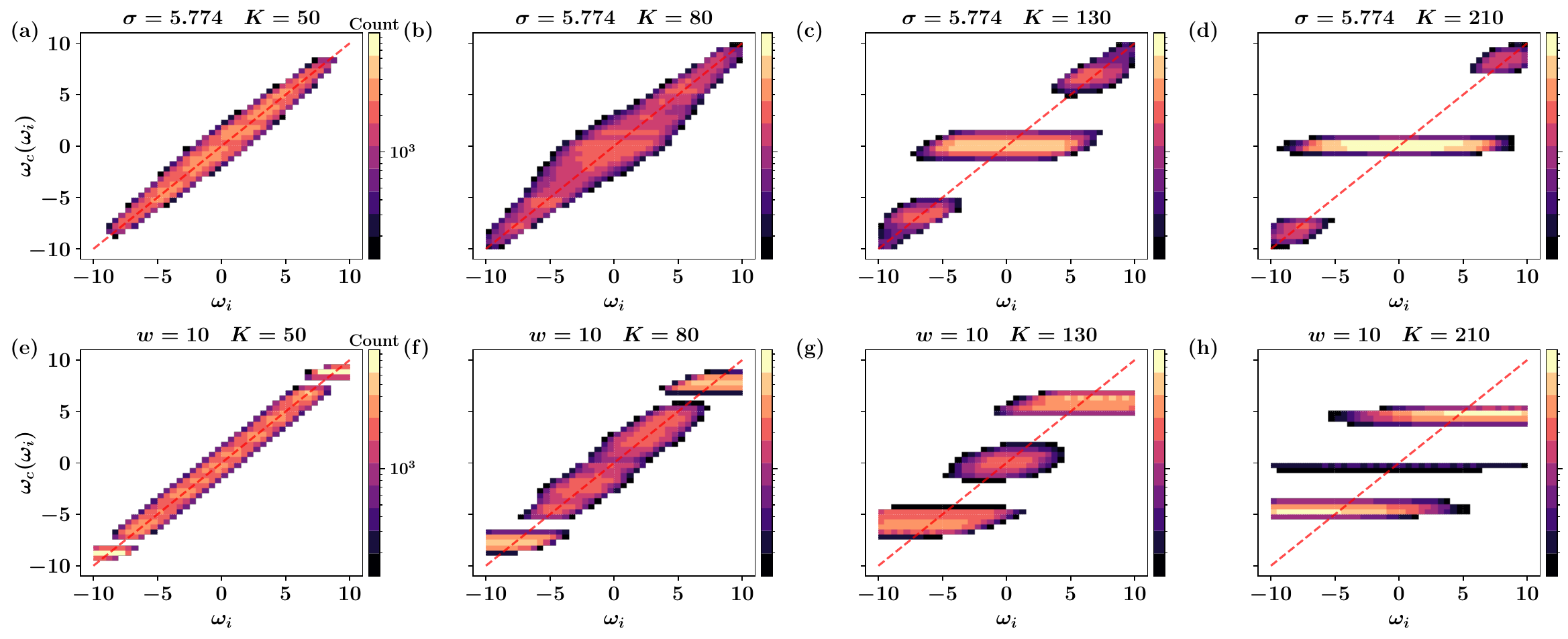}
\caption{
{\bf Breakdown of frequency-based cluster capture.}
Probability density of $(\omega_i,\omega_c)$ pairs across 512 realizations, relating each oscillator's intrinsic frequency $\omega_i$ to the mean frequency $\omega_c$ of the cluster it joins, for $m=10$.
(a--d) Gaussian $\sigma=5.774$ and (e--h) uniform $w=10$, at $K=50$, $80$, $130$, and $210$.
At low $K$, both distributions concentrate along the diagonal $\omega_i\approx\omega_c$, reflecting frequency-proximity capture.
As $K$ increases, the Gaussian retains a dominant diagonal core, whereas the uniform fractures into multiple horizontal and vertical bands.
Horizontal bands mean a single cluster recruits oscillators across a wide $\omega_i$ range; vertical bands mean oscillators of identical $\omega_i$ end up in different clusters depending on initial conditions---a direct signature of multistability.
}
\label{fig:rev3}
\end{figure}

First, Fig.~\ref{fig:rev3} relates each oscillator's intrinsic frequency $\omega_i$ to the mean frequency $\omega_c$ of the cluster it joins. If the criterion held, captured oscillators would lie along the diagonal $\omega_i\approx\omega_c$, and at weak coupling both distributions do [Fig.~\ref{fig:rev3}(a,e)]. As $K$ increases, the Gaussian keeps a dominant diagonal core, but the uniform fractures into multiple horizontal and vertical bands [Fig.~\ref{fig:rev3}(g,h)]: horizontal bands mean a single cluster recruits oscillators across a wide $\omega_i$ range, vertical bands mean oscillators of identical $\omega_i$ end up in different clusters depending on initial conditions---a direct signature of the multistability characterized in Sec.~\ref{sec:multistability}.

\begin{figure}[h]
\centering
\includegraphics[width=1.0\linewidth]{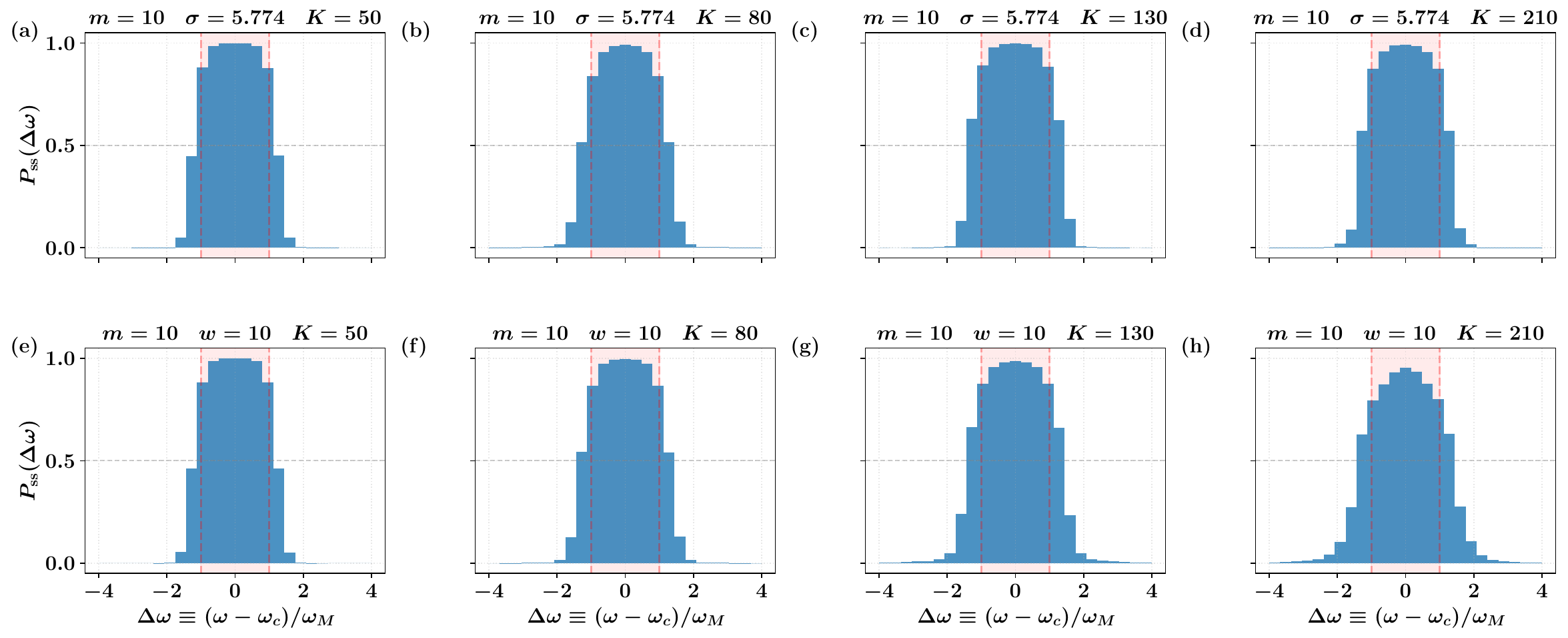}
\caption{
{\bf Steady-state capture probability and the failure of the Melnikov step.}
Steady-state capture probability $P_{\mathrm{ss}}(\Delta\omega)$ versus normalized frequency offset $\Delta\omega\equiv(\omega-\omega_c)/\omega_M$ for $m=10$.
(a--d) Gaussian $\sigma=5.774$ and (e--h) uniform $w=10$, at $K=50$, $80$, $130$, and $210$.
The single-cluster Melnikov prediction is a sharp step: unity for $|\Delta\omega|<1$ and zero outside.
At weak coupling the data follow this step, but as $K$ increases the uniform develops a smooth crossover, with substantial capture beyond $|\Delta\omega|=1$ and incomplete capture within it, signaling the breakdown of the single-cluster criterion under inter-cluster interaction.
}
\label{fig:rev7}
\end{figure}

Second, Fig.~\ref{fig:rev7} shows the steady-state capture probability $P_{\mathrm{ss}}$ versus normalized frequency offset $\Delta\omega\equiv(\omega-\omega_c)/\omega_M$. The single-cluster prediction is a sharp step---unity for $|\Delta\omega|<1$, zero outside---and the data follow it at weak coupling [Fig.~\ref{fig:rev7}(a,e)]. As $K$ increases, however, the uniform turns the step into a smooth crossover [Fig.~\ref{fig:rev7}(g,h)]: substantial capture persists beyond $\omega_M$, while oscillators inside are no longer guaranteed to be captured. Summarized over all $K$, the capture probabilities inside ($a$) and outside ($b$) the Melnikov radius [Fig.~\ref{fig:rev8}(a,e)] make this explicit: $a$ falls below unity and $b$ rises well above zero as the coupling grows.

Both diagnostics fail in the same regime, showing that frequency separation alone cannot determine cluster membership once clusters interact. At leading order, before the clusters interact, each one---Gaussian or uniform---grows on its own like an isolated primary cluster, well described by the Melnikov radius $\omega_M$; what fails is not the radius but the assumption that a single such cluster dominates. As the coupling grows and comparable clusters pass from isolated to interacting, this assumption breaks down---latest for the narrow, low-inertia Gaussian, whose central cluster keeps its dominance, and progressively earlier as the inertia or the variance grows and comparable clusters proliferate, for both shapes alike. The actual determinant is identified next.

\clearpage
\newpage


\section{Initial Energy as the Primary Determinant}
\label{sec:initial_energy}

\begin{figure}[h]
\centering
\includegraphics[width=1.0\linewidth]{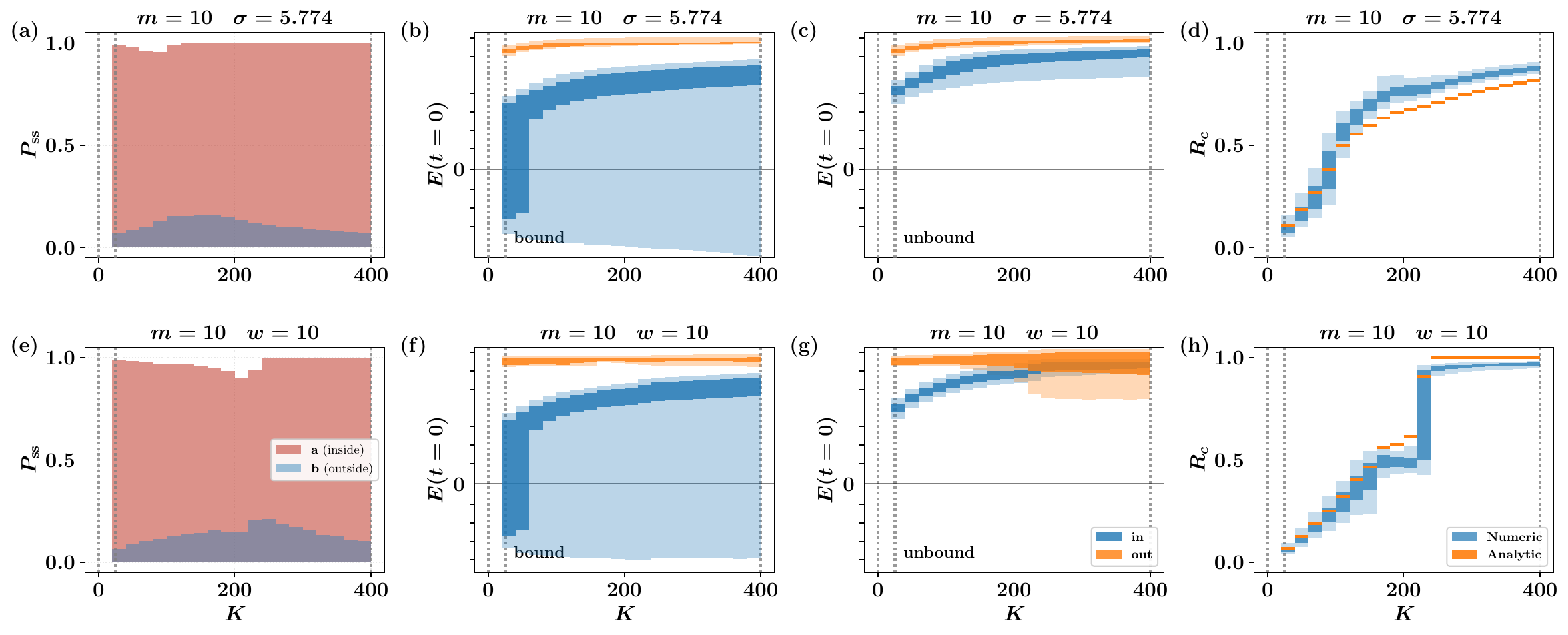}
\caption{
{\bf Energy-based capture and reduction to a probabilistic capture model.}
(a,e) Capture probabilities $a$ (for oscillators inside the Melnikov radius) and $b$ (outside) versus $K$ for Gaussian $\sigma=5.774$ and uniform $w=10$ ($m=10$); $a$ falls below unity and $b$ rises above zero as the coupling increases, making the failure of the single-cluster criterion explicit.
(b,f) Distribution of the initial energy $E(t{=}0)$ for oscillators that are eventually captured by a cluster and (c,g) for those that are not, shown separately for the inside and outside populations.
Captured oscillators carry systematically lower initial energy, identifying the initial energy as the primary capture determinant.
(d,h) Cluster order parameter $R_c$ versus $K$: numerical measurement (Numeric) compared with the prediction of the simplified capture model (Analytic), which uses only the probabilities $a$ and $b$.
The agreement confirms that the multistable dynamics reduce to a stochastic capture-and-escape process about the Melnikov radius, the energy-based selection entering only through $a$ and $b$.
}
\label{fig:rev8}
\end{figure}

The breakdown of frequency-based prediction is dynamical in origin. Computing the initial energy of each oscillator in its candidate cluster's center-of-mass frame,
\begin{align}
    E_i = \tfrac{1}{2} m \left( \dot{\theta}_{i} - \dot{\theta}_{c} \right)^{2} - K R_{c} \cos(\theta_{i} - \theta_{c}),
\end{align}
reveals a clear rule: for each cluster, the oscillators that are eventually captured carry systematically lower initial energy than those that are not [Fig.~\ref{fig:rev8}(b,c) and (f,g)]. The separation is robust inside the Melnikov radius at all $K$. Outside it, the two shapes diverge. For the Gaussian, whose dominant central cluster persists, the separation stays clean even at large $K$, captured and uncaptured oscillators remaining well sorted by initial energy. For the uniform case, it is weakened as $K$ grows: with no cluster dominating, peripheral oscillators are repeatedly exchanged between comparable clusters, so the two energy distributions overlap and their fate comes to depend on transient dynamics rather than initial energy alone. Even so, initial energy, not frequency, remains the primary capture determinant for the majority of oscillators in both cases, with the Melnikov criterion surviving only as a leading-order approximation.

The initial-condition sensitivity that defeats the mean-field description can itself be recast as a simple stochastic process. In both distributions, capture is governed by two probabilities read from Fig.~\ref{fig:rev8}(a,e): $a$, the probability inside the Melnikov radius ($|\omega-\omega_c|<\omega_M$), and $b$, the probability in the outer band ($\omega_M<|\omega-\omega_c|<4\omega_M$). The order parameter follows from $R_c=\int g(\omega)\,P_{\mathrm{capture}}(\omega)\,d\omega$, with $\omega_M=(4\gamma/\pi)\sqrt{KR_c/m}$ (dropping the subleading correction~\cite{gao2018self}, as $m$ and $K$ are large); the two shapes differ only in how the flat or peaked $g(\omega)$ weights this integral.

For the uniform distribution, $g(\omega)=1/(2w)$ is flat, so the capture probabilities multiply the integration window uniformly. Collecting the inner band (width $2\omega_M$, probability $a$) and the outer band (width $6\omega_M$, probability $b$) into a window-averaged probability $P\equiv(2a+6b)/8$,
\begin{align}
R_c \approx \frac{8\omega_M}{2w}\,P = \frac{4\omega_M}{w}\,P,
\end{align}
which, with $\omega_M=(4\gamma/\pi)\sqrt{KR_c/m}$, closes into
\begin{align}
R_c \approx \left(\frac{16\gamma P}{\pi w}\right)^{2}\frac{K}{m}.
\end{align}

For the Gaussian, $g(\omega)$ is peaked, so the inner and outer bands carry unequal mass and cannot be collapsed into a single window. We instead weight each band by its own Gaussian mass,
\begin{align}
R_c \approx a\,I_{\mathrm{in}} + b\,I_{\mathrm{out}},
\end{align}
where $I_{\mathrm{in}}$ and $I_{\mathrm{out}}$ are the mass inside the Melnikov radius and in the outer band,
\begin{align}
I_{\mathrm{in}}  &= \Phi\!\left(\frac{\omega_c+\omega_M}{\sigma}\right) - \Phi\!\left(\frac{\omega_c-\omega_M}{\sigma}\right), \\
I_{\mathrm{out}} &= \left[\Phi\!\left(\frac{\omega_c+4\omega_M}{\sigma}\right) - \Phi\!\left(\frac{\omega_c+\omega_M}{\sigma}\right)\right]
                  + \left[\Phi\!\left(\frac{\omega_c-\omega_M}{\sigma}\right) - \Phi\!\left(\frac{\omega_c-4\omega_M}{\sigma}\right)\right],
\end{align}
with $\Phi$ the standard normal CDF. For $a=b$ this collapses to $a\,(I_{\mathrm{in}}+I_{\mathrm{out}})$, recovering the uniform form.

Both relations are solved numerically for $R_c$ with the capture probabilities of Fig.~\ref{fig:rev8}(a,e), and the prediction (Analytic) matches the simulation (Numeric) throughout the transition [Fig.~\ref{fig:rev8}(d,h)]. The model uses only these two probabilities $a$ and $b$, with no reference to the initial energy of individual oscillators: once the energy-based selection is summarized in $a$ and $b$, the multistable, stepwise dynamics reduce to a stochastic capture-and-escape process about the Melnikov radius. This closes the quantitative picture: the multi-cluster dynamics break the single-cluster criterion, the variance and inertia set the degree of the resulting multistability, and all that remains of the microscopic complexity is the pair of probabilities $a$ and $b$.

\clearpage
\newpage


\section{Consistency of the Number Fractions Across Inertia and Variance}
\label{sec:RS5rho}

\begin{figure}[h]
\centering
\includegraphics[width=1.0\linewidth]{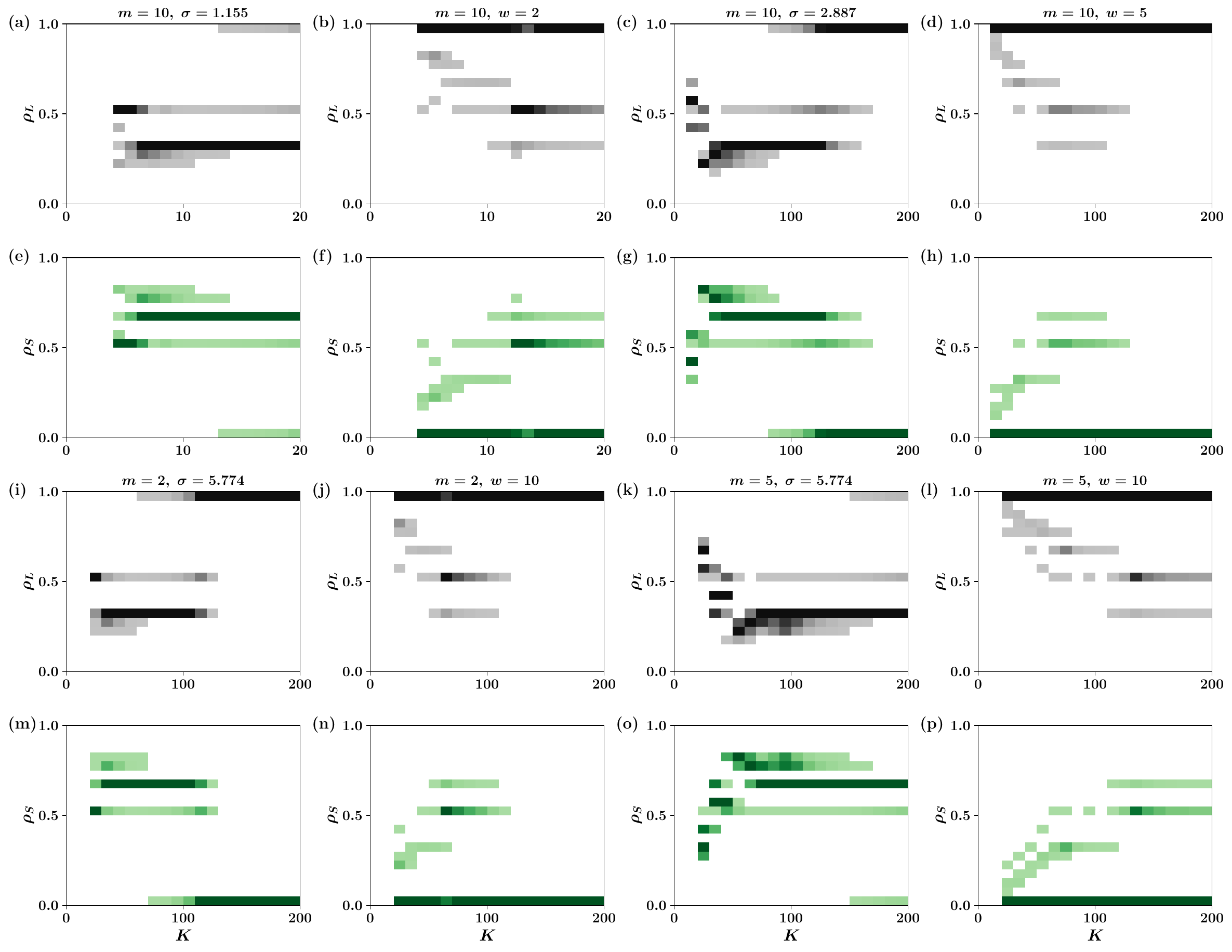}
\caption{
{\bf Cluster size structure: number fractions.}
Number fractions $\rho_L$, $\rho_S$ of large and small clusters ($\rho_L+\rho_S=1$) versus coupling strength $K$, across variance and inertia.
The upper half (a--h) fixes the inertia ($m=10$) and varies the variance (Gaussian $\sigma=1.155$, uniform $w=2$, Gaussian $\sigma=2.887$, uniform $w=5$); the lower half (i--p) fixes the variance ($\sigma=5.774$, $w=10$) and varies the inertia (Gaussian $m=2$, uniform $m=2$, Gaussian $m=5$, uniform $m=5$).
Within each half, the first row shows $\rho_L$ and the second $\rho_S$.
Each panel shows the ensemble distribution at each $K$ as a shaded histogram (darker $=$ more frequent).
At every variance and inertia, the Gaussian holds many small clusters ($\rho_L$ low, $\rho_S$ high) while the uniform is almost all large ($\rho_L\to1$, $\rho_S\to0$).
}
\label{fig:figRS5a}
\end{figure}

The number-fraction contrast that distinguishes the two shapes in the main text [Fig.~\ref{fig:rev5}] holds throughout parameter space. Figure~\ref{fig:figRS5a} spans it in two halves: the upper half varies the variance at fixed inertia ($m=10$), the lower half varies the inertia at matched variance.

The size structure splits the same way in every case: the Gaussian carries many small clusters alongside its dominant one ($\rho_L$ low, $\rho_S$ high), whereas the uniform is almost entirely large clusters ($\rho_L\to1$, $\rho_S\to0$). This size contrast is therefore a robust signature of the distribution shape rather than of any particular inertia or variance.

\clearpage
\newpage


\section{Consistency of the Devil's Staircase Statistics Across Inertia and Variance}
\label{sec:RS6}

\subsection{Intrinsic ratios across inertia and variance}

\begin{figure}[h]
\centering
\includegraphics[width=1.0\linewidth]{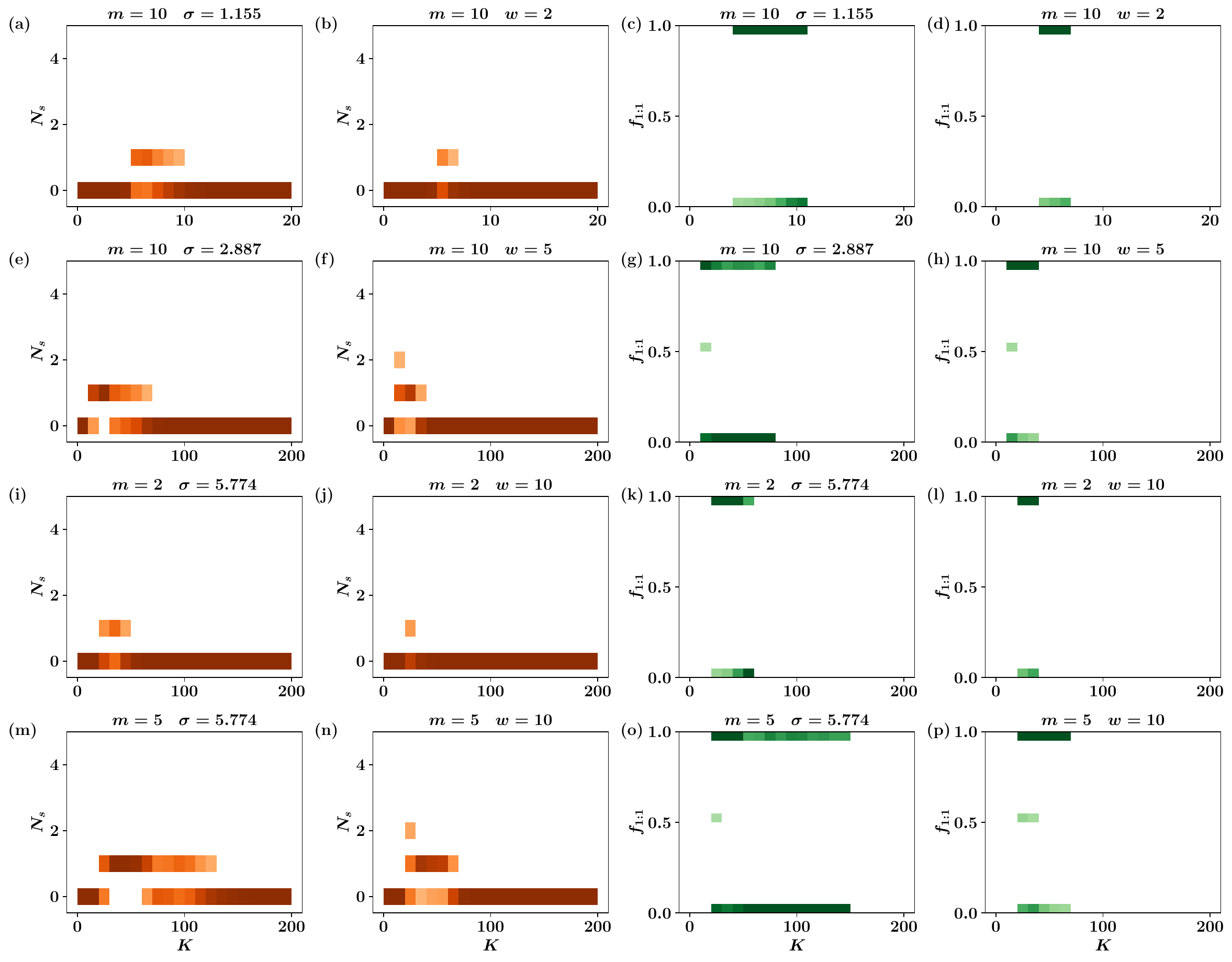}
\caption{
{\bf DS ratios across inertia and variance.}
Modal number of staircases $N_s$ [(a,b,e,f,i,j,m,n)] and equal-gap fraction $f_{1:1}$ [(c,d,g,h,k,l,o,p)] versus coupling strength $K$.
Within each group of four panels, the two left columns are Gaussian and the two right are uniform.
The upper half (a--h) fixes the inertia ($m=10$) and varies the variance (Gaussian $\sigma=1.155$, uniform $w=2$; Gaussian $\sigma=2.887$, uniform $w=5$); the lower half (i--p) fixes the variance ($\sigma=5.774$, $w=10$) and varies the inertia ($m=2$; $m=5$).
Each quantity is shown as a shaded histogram over the ensemble of realizations at each $K$ (darker $=$ more frequent), following the conventions of Fig.~\ref{fig:rev12}.
}
\label{fig:figRS6a}
\end{figure}

The main text shows that the distribution shape sets which rational ratios the staircase takes---hierarchical for the Gaussian, equal for the uniform. This contrast holds across parameter space [Fig.~\ref{fig:figRS6a}].

As $K$ grows, the modal number of staircases $N_s$ falls toward full locking at every inertia and variance. The equal-gap fraction $f_{1:1}$ selects which ratios appear, but only where enough clusters form for a hierarchy to develop. At larger inertia and variance the Gaussian develops the hierarchical ratios of its dominant central cluster, such as $2\!:\!1$, so $f_{1:1}$ is low for the Gaussian and high for the uniform, as in the main text [Fig.~\ref{fig:figRS6a}(o,p)].

At small inertia and variance the contrast narrows. Clusters of comparable size can no longer coexist across a broad frequency range, so few clusters form and the staircase shrinks to a single dominant cluster flanked by one small cluster on each side. This symmetric small--large--small arrangement has equal gaps about the center, so its two gaps appear as $1\!:\!1$ even though the organization is hierarchical, and $f_{1:1}$ rises for the Gaussian toward the high values of the uniform [Fig.~\ref{fig:figRS6a}(c,d)]. The hierarchical signature reappears only once the inertia and variance are large enough for several clusters to flank the center asymmetrically. The shape thus still governs \emph{which} ratios appear, provided the staircase carries enough clusters for them to differ. This distinction is a robust signature of the distribution shape rather than of any particular inertia or variance.

\subsection{Precision and participation of the locking}

\begin{figure}[h]
\centering
\includegraphics[width=1.0\linewidth]{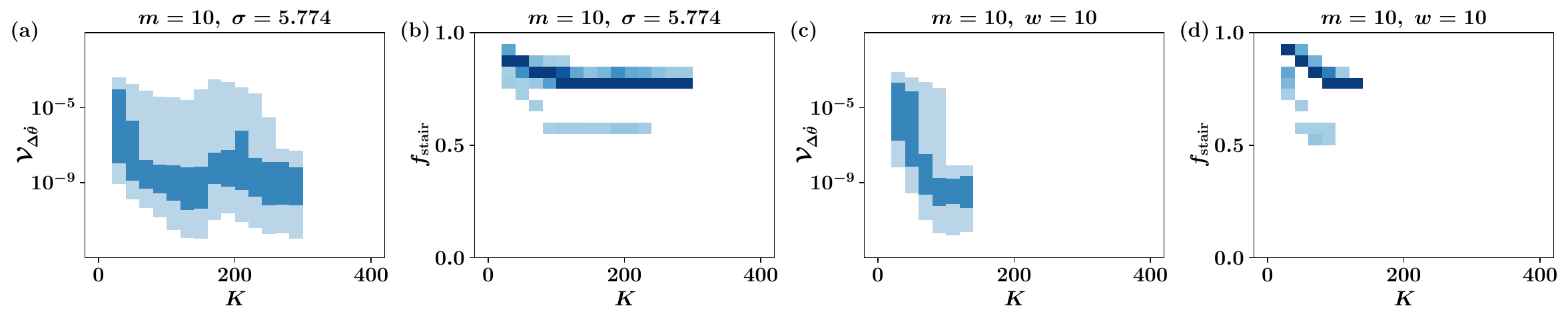}
\caption{
{\bf Precision and participation of the rational locking.}
Normalized gap variance $\mathcal{V}_{\Delta\dot\theta}$ [(a,c)] and staircase participation $f_{\mathrm{stair}}$ [(b,d)] versus coupling strength $K$ for $m=10$: (a,b) Gaussian $\sigma=5.774$ and (c,d) uniform $w=10$.
Panels (a,c) show the interquartile (dark) and $5$--$95\%$ (light) ranges of $\mathcal{V}_{\Delta\dot\theta}$ across realizations; panels (b,d) show the ensemble distribution of $f_{\mathrm{stair}}$ as a shaded histogram (darker $=$ more frequent).
For both shapes $\mathcal{V}_{\Delta\dot\theta}$ [Eq.~\eqref{eq:gapvar}] stays below $10^{-4}$ [(a,c)] and $f_{\mathrm{stair}}$ is near unity throughout [(b,d)], so the clusters lock into precise rational ratios and almost all of them participate in a staircase.
}
\label{fig:figS_RevS9}
\end{figure}

\begin{figure}[h]
\centering
\includegraphics[width=1.0\linewidth]{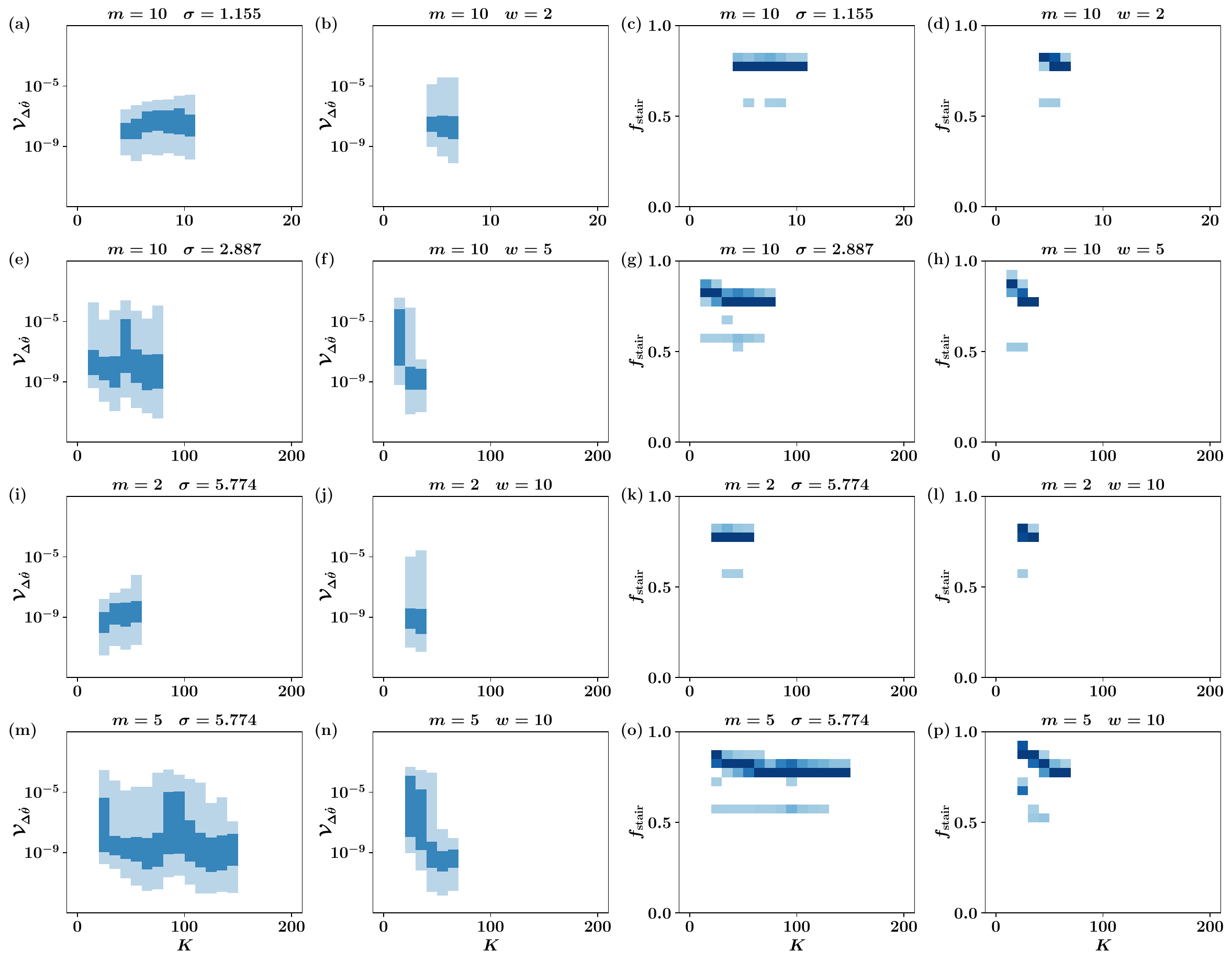}
\caption{
{\bf Precision and participation across inertia and variance.}
Normalized gap variance $\mathcal{V}_{\Delta\dot\theta}$ [(a,b,e,f,i,j,m,n)] and staircase participation $f_{\mathrm{stair}}$ [(c,d,g,h,k,l,o,p)] versus coupling strength $K$.
Within each group of four panels, the two left columns are Gaussian and the two right are uniform.
The upper half (a--h) fixes the inertia ($m=10$) and varies the variance (Gaussian $\sigma=1.155$, uniform $w=2$; Gaussian $\sigma=2.887$, uniform $w=5$); the lower half (i--p) fixes the variance ($\sigma=5.774$, $w=10$) and varies the inertia ($m=2$; $m=5$).
The $\mathcal{V}_{\Delta\dot\theta}$ panels show the interquartile (dark) and $5$--$95\%$ (light) ranges; the $f_{\mathrm{stair}}$ panels show the ensemble distribution as a shaded histogram (darker $=$ more frequent).
}
\label{fig:figRS6b}
\end{figure}

The ratios above presume that the clusters lock into a staircase precisely and that almost all of them take part. These two properties---the precision of the locking and the participation in it---underpin the staircase itself. We define them here and verify them at the representative parameters [Fig.~\ref{fig:figS_RevS9}], then across inertia and variance [Fig.~\ref{fig:figRS6b}].

Arranging a staircase's clusters by their angular velocities produces a sequence of adjacent gaps $\{\Delta\dot\theta_k\}_{k=1}^{n}$. The staircase is rationally locked when these gaps are integer multiples of a single base gap $g_0$, i.e.\ $\Delta\dot\theta_k \equiv p_k\,g_0$ with $p_k \in \mathbb{Z}$. In practice the locking is not exact; the ratios of adjacent gaps lie only close to rational numbers,
\begin{equation}
\frac{\Delta\dot\theta_{k+1}}{\Delta\dot\theta_k} \approx \frac{p_{k+1}}{p_k} \equiv q_{k,k+1} \in \mathbb{Q},
\end{equation}
from which we assign each gap its integer $p_k$, retaining only ratios that reduce to a denominator of at most $5$, the largest value we observe. A ratio is accepted as rational when this best approximation lies within $4\times10^{-2}$ of the measured value; otherwise the pair of gaps is deemed non-rational. A staircase is a maximal run of consecutive clusters whose successive gap ratios are all accepted as rational, and only runs of at least three clusters are counted; the number of such staircases in a realization is $N_s$. Dividing each gap by its integer,
\begin{equation}
\tilde g_k \equiv \Delta\dot\theta_k/p_k,
\end{equation}
reduces to a common base gap $g_0$ for every $k$ when the locking is exact, and spreads around it when it is not. The normalized gap variance
\begin{align}
\mathcal{V}_{\Delta\dot\theta} = \frac{\mathrm{Var}(\tilde g_k)}{\langle \tilde g_k\rangle^{2}},
~{\rm where}~\mathrm{Var}(\tilde g_k) = \frac{1}{n}\sum_{k=1}^{n}\bigl(\tilde g_k - \langle \tilde g_k\rangle\bigr)^{2},
\label{eq:gapvar}
\end{align}
is the squared coefficient of variation of the $\tilde g_k$ within a single staircase: it tends to zero as the $\tilde g_k$ converge to $g_0$ and grows as they spread. The staircase participation $f_{\mathrm{stair}}$ is the fraction of clusters that belong to a staircase.

At $m=10$, the locking is precise for both shapes: $\mathcal{V}_{\Delta\dot\theta}$ stays below $10^{-4}$ [Fig.~\ref{fig:figS_RevS9}(a,c)], so the angular-velocity gaps lie on clean rational ratios. The participation is likewise high, $f_{\mathrm{stair}}$ remaining near unity throughout [Fig.~\ref{fig:figS_RevS9}(b,d)], so almost every cluster joins a staircase. This high participation is consistent with rational locking being energetically favorable, the commensurate ratios acting as the preferred configuration rather than a rare coincidence.

Both properties are universal. At every inertia and variance $\mathcal{V}_{\Delta\dot\theta}$ stays below $10^{-4}$ and $f_{\mathrm{stair}}$ remains near unity [Fig.~\ref{fig:figRS6b}], so the clusters lock precisely and almost all of them participate, regardless of shape. The precision and participation of the staircase are therefore fixed features of the dynamics, independent of the distribution shape---the stable ground on which the shape sets only \emph{which} ratios appear.

\clearpage
\newpage


\section{Consistency of the Seeded Fractions Across Inertia and Variance}
\label{sec:RS5f}

\begin{figure}[h]
\centering
\includegraphics[width=1.0\linewidth]{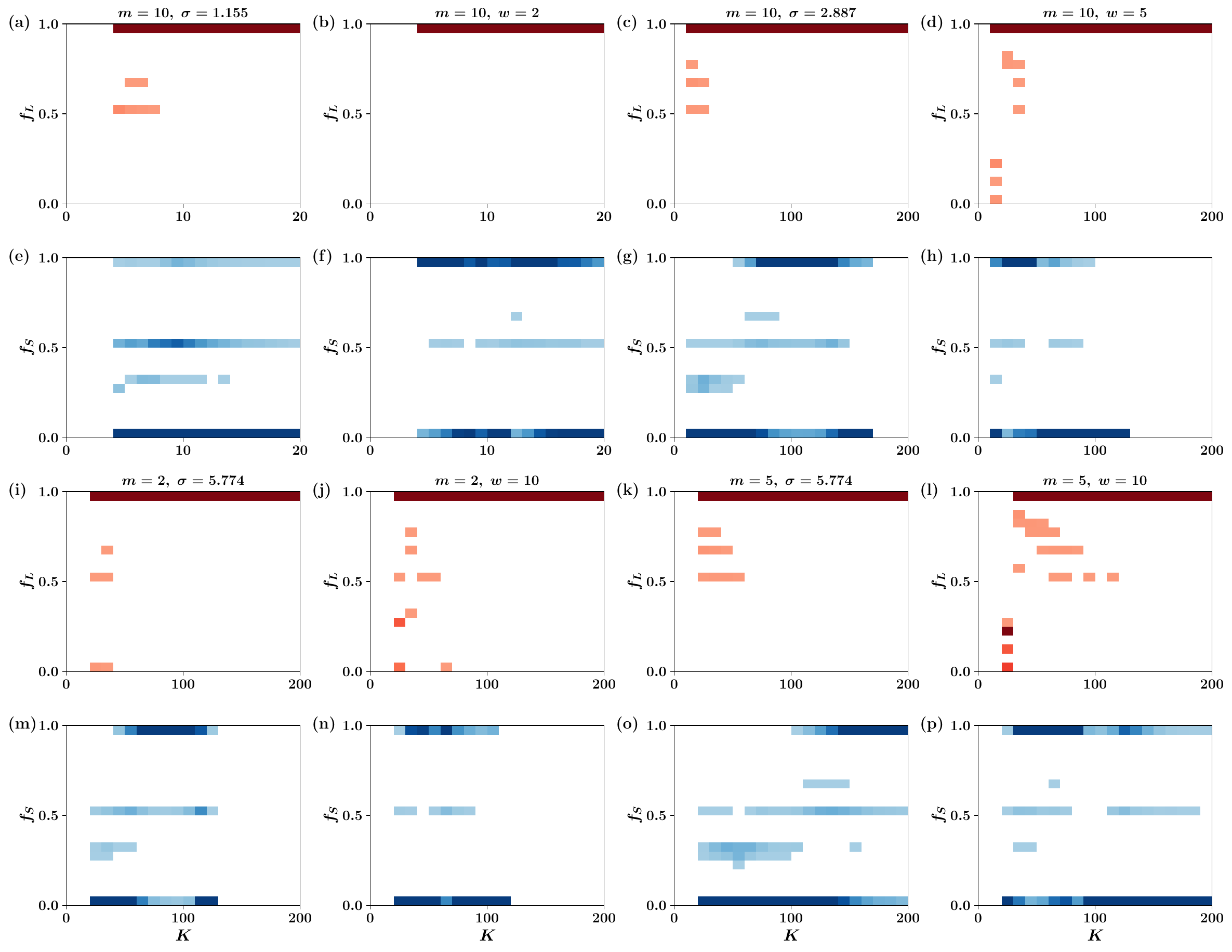}
\caption{
{\bf Cluster size structure: seeded fractions.}
Seeded fractions $f_L$, $f_S$ of large and small clusters versus coupling strength $K$, with the same layout as Fig.~\ref{fig:figRS5a}: the upper half (a--h) varies the variance at fixed inertia ($m=10$), the lower half (i--p) varies the inertia at matched variance, and within each half the first row shows $f_L$ and the second $f_S$.
Across parameter space, $f_L$ stays high for both shapes, whereas $f_S$ varies more intricately with coupling, inertia, and variance.
}
\label{fig:figRS5b}
\end{figure}

The seeded fractions report when each cluster is established relative to the seed stage, and here the large and small clusters diverge [Fig.~\ref{fig:figRS5b}]: $f_L$ is robust across parameter space, while $f_S$ is not.

The large clusters are robust. At every inertia and variance $f_L$ stays high for both shapes, so the large clusters are already in place at the seed stage regardless of the distribution. This confirms the ordering assumed in the main text---the large clusters form first, the small ones only afterward.

The small clusters are more intricate. At the matched pair the main text reports a clean contrast, $f_S$ rising with $K$ for the Gaussian and falling for the uniform, but this picture softens across the full parameter range. The trend persists only broadly: $f_S$ no longer follows a single monotonic rule, its coupling dependence reshaping with inertia and variance as the independent nucleation of small clusters competes with their capture by the growing dominant cluster. The small-cluster timing is thus shape-dependent but too intricate to reduce to one universal trend, unlike the uniformly high $f_L$ of the large clusters.

\clearpage
\newpage


\section{Consistency of the Seeding Migration Across Inertia and Variance}
\label{sec:RS8}

\begin{figure}[h]
\centering
\includegraphics[width=1.0\linewidth]{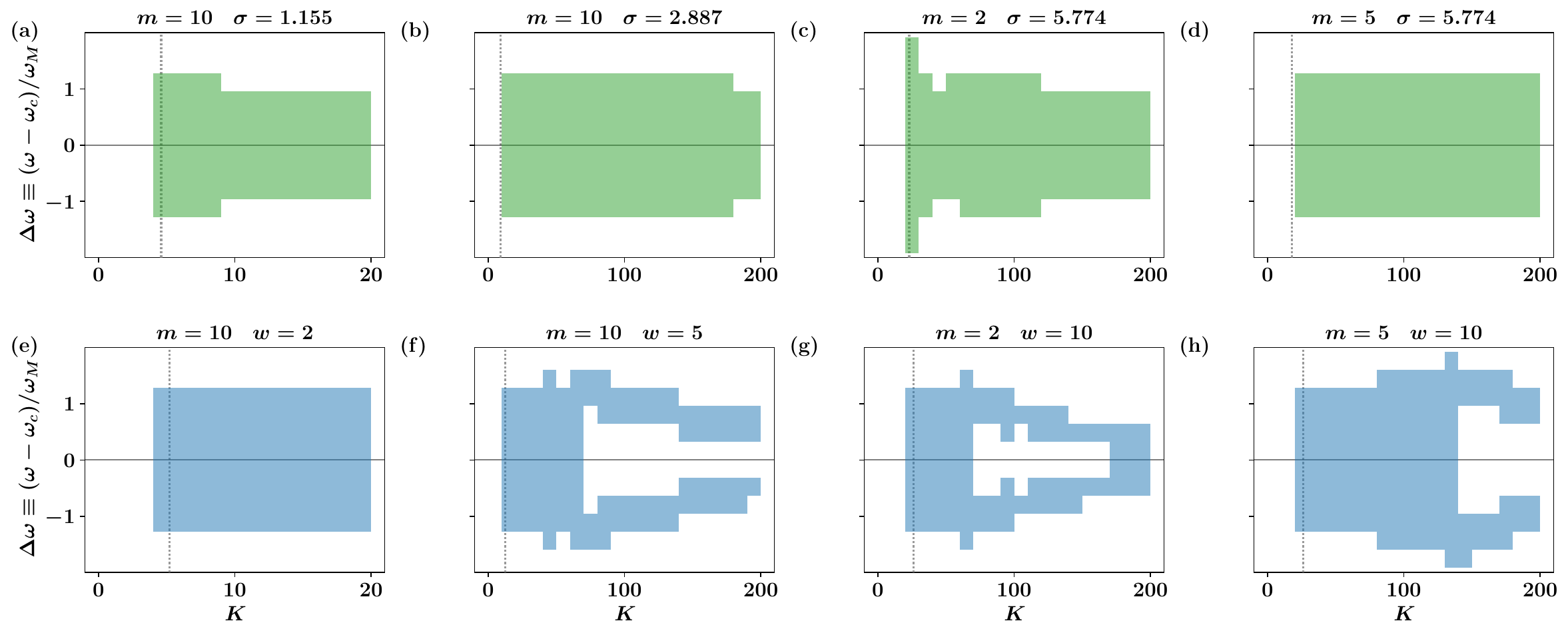}
\caption{
{\bf Seeding migration across inertia and variance.}
High-$P_{\mathrm{ts}}$ region in the normalized frequency offset $\Delta\omega\equiv(\omega-\omega_c)/\omega_M$ versus $K$, following the convention of Fig.~\ref{fig:rev10}: at each $K$, shaded bars mark the $\Delta\omega$ bins where the transient capture probability $P_{\mathrm{ts}}$ of the large clusters exceeds its mean, and the grey dotted line is the critical coupling.
(a--d) Gaussian and (e--h) uniform, at $m=10$ with $\sigma=1.155$ ($w=2$) and $\sigma=2.887$ ($w=5$), and at matched variance $\sigma=5.774$ ($w=10$) with $m=2$ and $m=5$.
For the Gaussian the high-$P_{\mathrm{ts}}$ region stays within the Melnikov region ($|\Delta\omega|\lesssim1$) throughout, whereas for the uniform it develops a central dip and migrates beyond the Melnikov boundary as $K$ grows.
}
\label{fig:figRS8}
\end{figure}

The seeding migration that distinguishes the two shapes in the main text [Fig.~\ref{fig:rev10}] holds across parameter space. Figure~\ref{fig:figRS8} varies both the variance (at fixed $m=10$) and the inertia (at matched variance).

For the Gaussian the high-$P_{\mathrm{ts}}$ region stays within the Melnikov region ($|\Delta\omega|\lesssim1$) at every inertia and variance [Fig.~\ref{fig:figRS8}(a--d)], its dense central peak seeding the giant cluster at the center. For the uniform case, the region instead develops a central dip and splits into two lobes that migrate beyond $|\Delta\omega|=1$ as $K$ grows [Fig.~\ref{fig:figRS8}(e--h)], the signature of seeds pushed outward to the high-frequency periphery.

Inertia and variance set only the degree of this migration---most pronounced for the widest, highest-inertia cases---while the shape fixes its direction: central for the Gaussian, central-to-peripheral for the uniform. The shift from entrainment- to merger-driven seeding is therefore a robust signature of the distribution shape, not of any particular inertia or variance.

\clearpage
\newpage


\section{Consistency of Pathway-Dependent Robustness Across Inertia and Variance}
\label{sec:RS9}

\begin{figure}[h]
\centering
\includegraphics[width=1.0\linewidth]{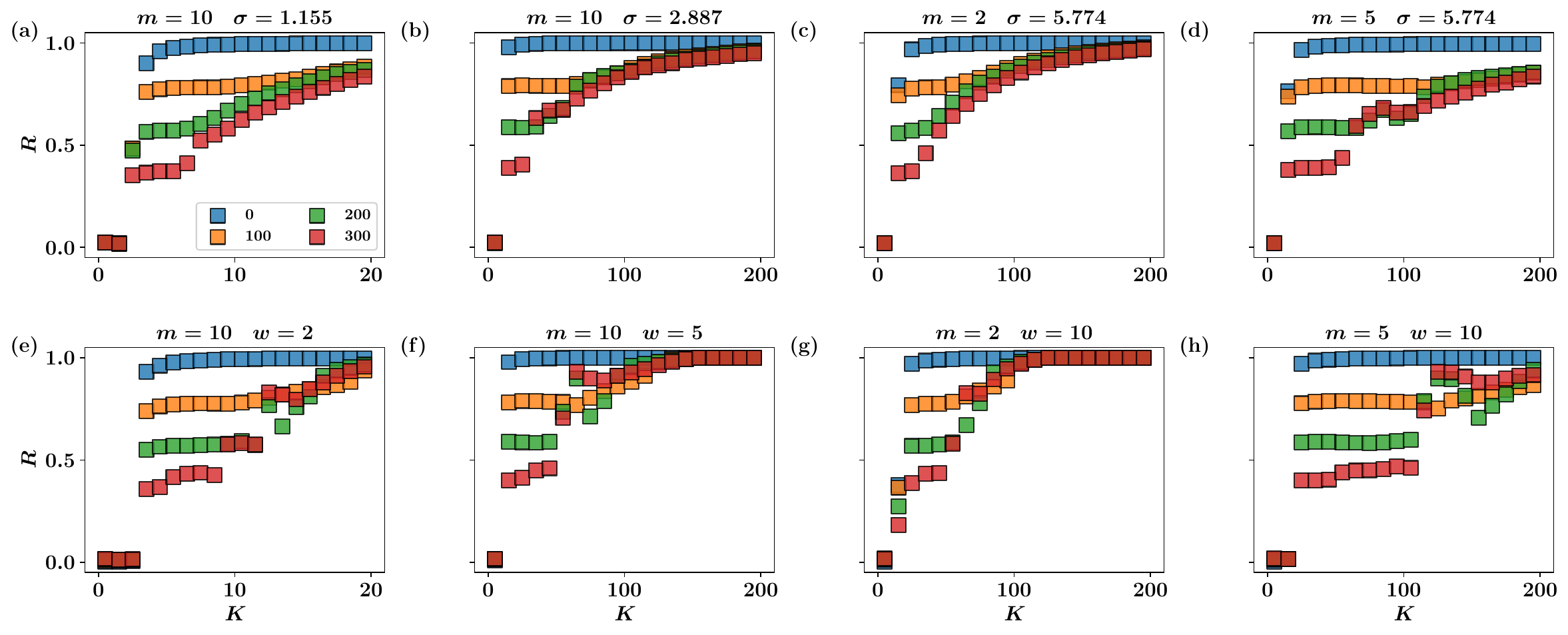}
\caption{
{\bf Pathway-dependent robustness across inertia and variance.}
Median order parameter $R$ versus coupling strength $K$ under peripheral perturbations of size $s=0$, $100$, $200$, and $300$, following the protocol of Fig.~\ref{fig:rev11}: the system is prepared in a maximally synchronized state by a backward process, and the $s$ lowest and $s$ highest intrinsic-frequency oscillators are perturbed by resetting their velocities to $\dot\theta_i=\omega_i/\gamma$ and randomizing their phases.
(a--d) Gaussian and (e--h) uniform, at $m=10$ with $\sigma=1.155$ ($w=2$) and $\sigma=2.887$ ($w=5$), and at matched variance $\sigma=5.774$ ($w=10$) with $m=2$ and $m=5$.
At every inertia and variance the Gaussian recovers readily at weak to intermediate $K$, whereas the uniform stays trapped in a low-$R$ plateau there and recovers only at strong $K$, where the perturbed--unperturbed gap closes.
}
\label{fig:figRS9}
\end{figure}

The pathway-dependent robustness that distinguishes the two shapes in the main text [Fig.~\ref{fig:rev11}] holds across parameter space. Figure~\ref{fig:figRS9} varies both the variance (at fixed $m=10$) and the inertia (at matched variance).

At every inertia and variance the two shapes recover in opposite coupling ranges. The Gaussian recovers readily at weak to intermediate $K$, its perturbed curves tracking the unperturbed one closely [Fig.~\ref{fig:figRS9}(a--d)], as its periphery is reclaimed oscillator by oscillator through re-entrainment. The uniform instead stays trapped in a low-$R$ plateau there, a wide gap separating its perturbed curves from the unperturbed one, and recovers only at strong $K$ [Fig.~\ref{fig:figRS9}(e--h)], where its displaced periphery re-forms clusters and rejoins by merger.

Inertia and variance set only the depth of the vulnerable plateau---most pronounced for the widest, highest-inertia cases---while the shape fixes the opposite-range pattern itself. The inheritance of stability from the assembly pathway is therefore a robust signature of the distribution shape, not of any particular inertia or variance.

\clearpage
\newpage




\renewcommand{\refname}{Supplementary References}
\putbib
\end{bibunit}

\end{document}